\documentclass[twocolumn]{aastex631}   
\usepackage{natbib}
\usepackage{xspace}

\newcommand{\be}{\begin{equation}}
\newcommand{\ee}{\end{equation}}

\newcommand{\Fermi}{\textit{Fermi}\xspace}
\newcommand{\Fermilat}{\textit{Fermi}-LAT\xspace}

\shorttitle{4FGL DR3}
\shortauthors{\Fermilat collaboration}

\begin{document}

 

\title{Incremental \Fermi Large Area Telescope Fourth Source Catalog}
    

\author{S.~Abdollahi}
\affiliation{IRAP, Universit\'e de Toulouse, CNRS, UPS, CNES, F-31028 Toulouse, France}
\author{F.~Acero}
\affiliation{Universit\'e Paris Saclay and Universit\'e Paris Cit\'e, CEA, CNRS, AIM, F-91191 Gif-sur-Yvette, France}
\author{L.~Baldini}
\affiliation{Universit\`a di Pisa and Istituto Nazionale di Fisica Nucleare, Sezione di Pisa I-56127 Pisa, Italy}
\author{J.~Ballet}
\email{jean.ballet@cea.fr}
\affiliation{Universit\'e Paris Saclay and Universit\'e Paris Cit\'e, CEA, CNRS, AIM, F-91191 Gif-sur-Yvette, France}
\author{D.~Bastieri}
\affiliation{Istituto Nazionale di Fisica Nucleare, Sezione di Padova, I-35131 Padova, Italy}
\affiliation{Dipartimento di Fisica e Astronomia ``G. Galilei'', Universit\`a di Padova, I-35131 Padova, Italy}
\affiliation{Center for Space Studies and Activities ``G. Colombo", University of Padova, Via Venezia 15, I-35131 Padova, Italy}
\author{R.~Bellazzini}
\affiliation{Istituto Nazionale di Fisica Nucleare, Sezione di Pisa, I-56127 Pisa, Italy}
\author{B.~Berenji}
\affiliation{California State University, Los Angeles, Department of Physics and Astronomy, Los Angeles, CA 90032, USA}
\author{A.~Berretta}
\affiliation{Dipartimento di Fisica, Universit\`a degli Studi di Perugia, I-06123 Perugia, Italy}
\author{E.~Bissaldi}
\affiliation{Dipartimento di Fisica ``M. Merlin" dell'Universit\`a e del Politecnico di Bari, via Amendola 173, I-70126 Bari, Italy}
\affiliation{Istituto Nazionale di Fisica Nucleare, Sezione di Bari, I-70126 Bari, Italy}
\author{R.~D.~Blandford}
\affiliation{W. W. Hansen Experimental Physics Laboratory, Kavli Institute for Particle Astrophysics and Cosmology, Department of Physics and SLAC National Accelerator Laboratory, Stanford University, Stanford, CA 94305, USA}
% ...


\begin{abstract}
We present an incremental version (4FGL-DR3, for Data Release 3) of the fourth \Fermilat catalog of $\gamma$-ray sources.
Based on the first twelve years of science data in the energy range from 50~MeV to 1~TeV, it contains 6658 sources. The analysis improves on that used for the 4FGL catalog over eight years of data: more sources are fit with curved spectra, we introduce a more robust spectral parameterization for pulsars, and we extend the spectral points to 1 TeV. The spectral parameters, spectral energy distributions, and associations are updated for all sources. Light curves are rebuilt for all sources with 1 yr intervals (not 2 month intervals).

Among the 5064 original 4FGL sources, 16 were deleted, 112 are formally below the detection threshold over 12 yr (but are kept in the list), while 74 are newly associated, 10 have an improved association, and seven associations were withdrawn. Pulsars are split explicitly between young and millisecond pulsars. Pulsars and binaries newly detected in LAT sources, as well as more than 100 newly classified blazars, are reported. We add three extended sources and 1607 new point sources, mostly just above the detection threshold, among which eight are considered identified, and 699 have a plausible counterpart at other wavelengths.

We discuss the degree-scale residuals to the global sky model and clusters of soft unassociated point sources close to the Galactic plane, which are possibly related to limitations of the interstellar emission model and missing extended sources.
\end{abstract}


\section{Introduction}
\label{introduction}

The \Fermi Large Area Telescope (LAT) has been surveying the high-energy $\gamma$-ray sky since 2008 \citep{LAT09_instrument}, and the LAT Collaboration has published a succession of source catalogs based on comprehensive analyses of LAT data.
The fourth source catalog (4FGL, where FGL stands for \Fermi Gamma-ray LAT) was derived from the analysis of the first 8 yr of LAT science data and contained 5064 sources \citep[][hereafter the 4FGL paper, or simply 4FGL]{LAT20_4FGL}. Every FGL catalog until and including 4FGL had used a new analysis method, new calibrations, a new diffuse model, and even a new reconstruction of the events themselves (like Pass 8 for 4FGL). Therefore each successive version of the catalog was largely independent of the previous one.
Results for different versions were compared only after the fact.

Since the development of the 4FGL, the data have been stable, no new model for the Galactic interstellar emission was built, and the analysis method has not evolved much. So we decided to change our approach and start building incremental 4FGL versions every two years, until one major analysis or data component changes. The Data Release 2 \citep[DR2;][]{LAT20_4FGLDR2} covered 10 yr of data and contained 723 additional sources. We refer to the original 4FGL sources as Data Release 1 (DR1).

In this paper, we describe the incremental procedure and apply it to 12 yr of data, creating the Data Release 3 (4FGL-DR3 or DR3 for short). The 4FGL paper remains the reference for the detailed methodology.
DR2 was strictly incremental (all DR1 sources are included and the analysis procedure was exactly the same). Meanwhile we have improved a few analysis steps, and recent publications have revealed a few new extended sources. Therefore we have relaxed this condition in DR3, to allow for deleting a few sources, and accept a few improved localizations.

All the information present in the DR1 catalog is updated, except the 2 month light curves. Those are very costly (in terms of CPU and disk space), and we have shown in the 4FGL paper that the 1 yr light curves capture most of the variability information. Monthly light curves, including indications of spectral variability, are now provided with similar accuracy in the \Fermilat Light Curve Repository\footnote{See \url{https://fermi.gsfc.nasa.gov/ssc/data/access/lat/LightCurveRepository/}.}. Detection of transient sources on monthly time scales is covered by 1FLT\footnote{See \url{https://www.ssdc.asi.it/fermi_iflt}.} \citep{2021_1FLT}. Transients on weekly time scales are covered by FAVA\footnote{See \url{https://fermi.gsfc.nasa.gov/ssc/data/access/lat/FAVA/}.} \citep{LAT17_FAVA}. Phase-folded spectral analysis of periodic sources (pulsars and binaries) is provided in dedicated papers.

The condition for switching the spectral models to a curved representation is now less stringent, thereby increasing the fraction of curved spectra to about half. We improved the spectral parameterization of pulsars, quantified the residuals using a new method, and now include in the Catalog FITS file the peak energy (in $\nu F_\nu$) for each source.

The community has worked hard on the unassociated LAT sources in recent years. Optical observations allowed for classifying many blazars, and radio, optical, and $\gamma$-ray timing allowed for detecting new pulsars and binary systems. We have included all this information in the source associations. The DR2 and DR3 results provide about 900 more unassociated sources to look into. Unassociated sources close to the Galactic plane are spatially clustered and have distinct spectral characteristics. We have singled them out by a specific flag.

Section~\ref{lat_and_background} describes the data and the updates to the diffuse model, Section~\ref{catalog_main} describes the updates to the analysis including the new extended sources,
Section~\ref{dr3_description} describes the results including a comparison to DR1, Section~\ref{dr3_assocs} describes the updates to the associations, and Section~\ref{dr3_unassocs} discusses the unassociated sources, with particular emphasis on the category of soft clustered sources close to the Galactic plane.

\section{Instrument and Background}
\label{lat_and_background}

\subsection{The LAT Data}
\label{LATData}

The data for the 4FGL-DR3 catalog were taken during the period 2008 August 4 (15:43 UTC) to 2020 August 2 (8:33 UTC) covering twelve years.
During most of this time, \Fermi was operated in sky-scanning survey mode, with the viewing direction rocking north and south of the zenith on alternate orbits such that the entire sky is observed every $\sim$3 hr.
Starting on 2018 March 16, the \Fermi spacecraft was put in safe hold after one solar array drive became stuck. Scientific operations of the LAT were interrupted for more than three weeks, by far the longest missing time interval since 2008, and restarted on April 8 in partial sky-scanning mode\footnote{See \url{https://fermi.gsfc.nasa.gov/ssc/observations/types/post_anomaly/}.}. The Sun more rarely enters the field of view, and during some phases of the $\sim$53-day precession period of the orbit, the entire sky is not covered every three hours. This hampers the solar observations and short-term light curves, but has little impact on the integrated sky coverage.

\begin{deluxetable*}{lrrrlllll}
\tablecaption{4FGL-DR3 Summed Likelihood components
\label{tab:components}
}
\tablehead{
\colhead{Energy Interval} & \colhead{NBins} & \colhead{ZMax} & \colhead{Ring Width} & \multicolumn{5}{c}{Pixel Size (deg)} \\
(GeV) &  & (deg) & (deg) & \colhead{PSF0} & \colhead{PSF1} & \colhead{PSF2} & \colhead{PSF3} & \colhead{All}
}
\startdata
0.05 -- 0.1 & 6 & 80 & 7 & \nodata & \nodata & \nodata & 0.6 & \nodata \\
0.1 -- 0.3 & 5 & 90 & 7 & \nodata & \nodata & 0.6 & 0.6 & \nodata \\
0.3 -- 1 & 6 & 100 & 5 & \nodata & 0.4 & 0.3 & 0.2 & \nodata \\
1 -- 3 & 5 & 105 & 4 & 0.4 & 0.15 & 0.1 & 0.1 & \nodata \\
3 -- 10 & 6 & 105 & 3 & 0.25 & 0.1 & 0.05 & 0.04 & \nodata \\
10 -- 30 & 5 & 105 & 2 & 0.15 & 0.06 & 0.04 & 0.03 & \nodata \\
30 -- 1000 & 8 & 105 & 1 & \nodata & \nodata & \nodata & \nodata & 0.03 \\
\enddata
\tablecomments{We used 19 components (all in binned mode) in the Summed Likelihood approach (\S~\ref{catalog_significance}) for DR3. Components in a given energy interval share the same number of energy bins, the same zenith angle selection, and the same region of interest (RoI) core, but have different pixel sizes in order to adapt to the PSF width. Each filled entry under Pixel Size corresponds to one component of the summed log-likelihood. NBins is the number of energy bins in the interval, ZMax is the zenith angle cut, Ring Width refers to the difference between the RoI core and the extraction region \citep[see][for details]{LAT20_4FGL}.
}
\end{deluxetable*}

As in 4FGL, intervals around solar flares and bright $\gamma$-ray bursts (GRBs) were excised. During the additional four years, 108 ks were cut due to two successive bright solar flares in 2017 September, and 2.4 ks around 5 new bright GRBs.
The current version of the LAT data is Pass 8 P8R3 \citep{LAT13_P8, LAT18_P305}.
For DR2 and DR3, we used the P8R3\_V3 IRFs instead of P8R3\_V2 used to build DR1. The relative effective areas for the point-spread function (PSF) event types were recalibrated on bright sources\footnote{See \url{https://fermi.gsfc.nasa.gov/ssc/data/analysis/LAT_caveats_p8r3v3.html}.}, reducing the systematics from 20\% to 5\%.
The energy range remains 50~MeV to 1~TeV. In 4FGL, the event types were not treated separately above 10~GeV. We now decompose the events into 4 components up to 30~GeV, resulting in 19 components in total (Table~\ref{tab:components}), for better precision but at the cost of more disk space and CPU.

\subsection{Model for the Diffuse Gamma-Ray Background}
\label{DiffuseModel}

\begin{figure}[!ht]
\centering
\includegraphics[width=\linewidth]{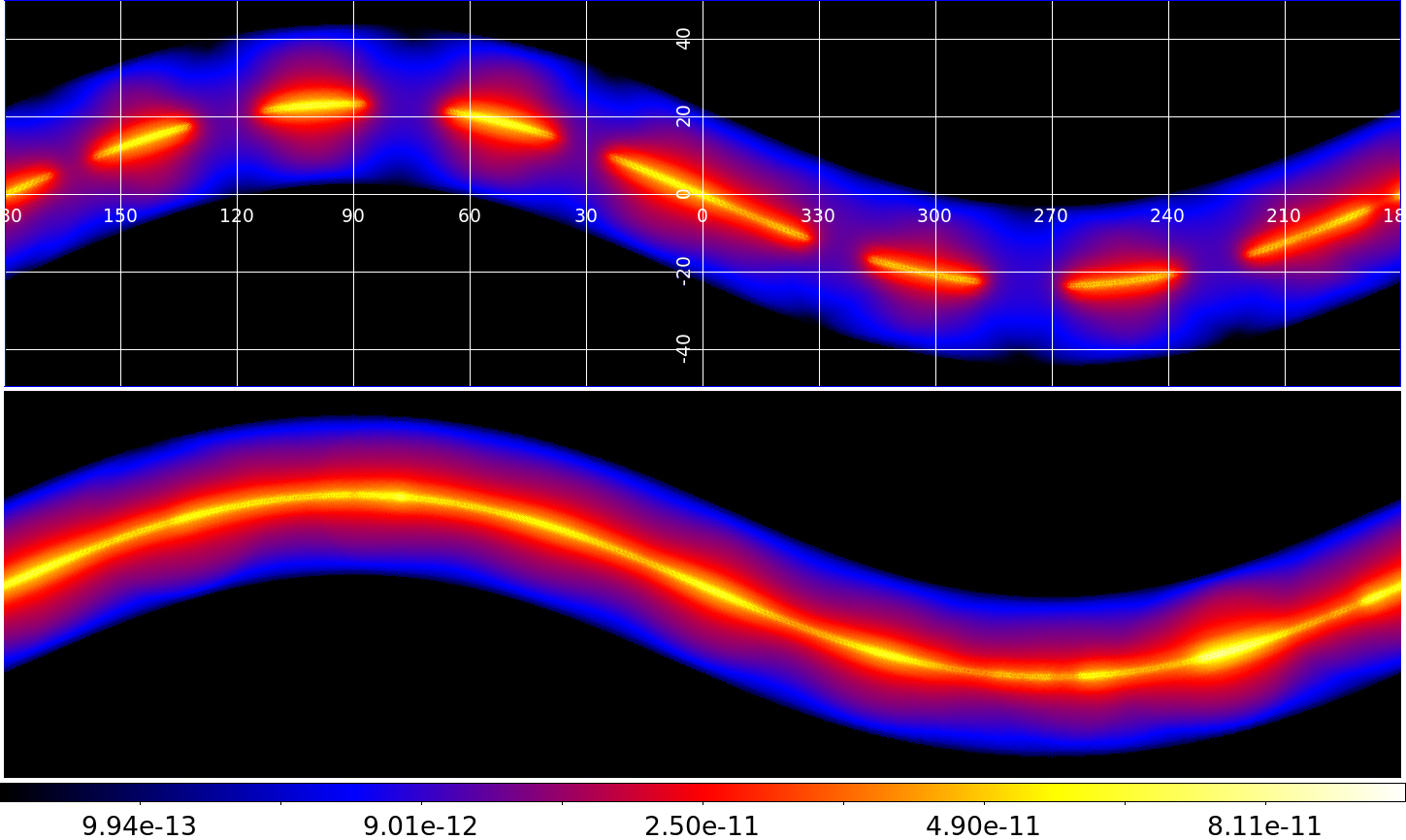}
\caption{Model for the inverse Compton emission of the Sun at 1 GeV averaged over one year, in equatorial coordinates. The top and bottom rows are, respectively, years 12 (2019--2020) and 9 (2016--2017), chosen to be after and before the solar array drive anomaly. The color bar is in sqrt scale, and the intensity unit is ph cm$^{-2}$ s$^{-1}$ sr$^{-1}$ MeV$^{-1}$.
}
\label{fig:sunic}
\end{figure}

The interstellar emission is a very important ingredient when characterizing the point sources, because it is structured and 90\% of the photons in the Galactic plane are of diffuse origin.
We used exactly the same model as in 4FGL\footnote{See \url{https://fermi.gsfc.nasa.gov/ssc/data/access/lat/BackgroundModels.html}.}. The isotropic spectrum was updated to P8R3\_V3.
The model for the emission of the Sun and the Moon was also kept the same, neglecting the modulation of the emission along the solar cycle. It became necessary, however, to generate a different effective model for each year (for the light curves). After the solar array drive anomaly during the tenth year, the coverage of the Sun changed significantly (Figure~\ref{fig:sunic}). For intervals of about two weeks, the Sun does not enter the field of view at all.

\section{Construction of the Catalog}
\label{catalog_main}

Most of the steps were identical to 4FGL \citep{LAT20_4FGL}.
As before the extended sources are taken from the literature. We only search for point sources.
We use the same Test Statistic (TS), TS = 2 $\ln (\mathcal{L} / \mathcal{L}_0)$, to quantify how significantly a source emerges from the background, comparing the maximum value of the likelihood function $\mathcal{L}$ including the source in the model with $\mathcal{L}_0$, the value without the source.
The 1 yr light curves were generated in exactly the same way.

\subsection{Extended Sources}
\label{catalog_extended}

Several extended sources (existing and new) were the subject of publications since the 4FGL paper. They were not considered in DR2, but were considered for DR3:
\begin{itemize}
  \item We adopted the Gaussian model proposed by \citet{G150_Devin20} for the broad supernova remnant (SNR) G150.3+4.5, after we confirmed that it is a better fit than the disk model used in DR1.
  \item \citet{VelaX_Tibaldo18} have studied the Vela X pulsar wind nebula (PWN) and reported that the radio template was superior to the geometric templates. We found that it does not provide a better fit than the DR1 model, but it eliminates six dubious point sources, so it is simpler and we adopted it. The same authors reported a spatial evolution as a function of energy, but we found that their high-energy disk is not significant when using the radio template.
  \item \citet{G279_Araya20} has shown that the cluster of DR1 sources around the SNR G279.0+1.1 is actually an extended LAT source. We adopted their ring model.
  \item HESS J1640$-$465 is seen as extended by H.E.S.S., but was considered a point source in the DR1 catalog. We adopted the H.E.S.S. Gaussian template, after it was shown by \citet{HESSJ1640_Mares21} that it is a better fit to the LAT data as well.
  \item Two papers discussed the \Fermilat emission from the PWN HESS J1825$-$137. The morphology considered by \citet{HESSJ1825_Araya19} was too complex for our catalog. We tried the broader Gaussian proposed by \citet{HESSJ1825_Principe20} but found it worsened the fit. We did change the template, however, because we noticed that the one that we used since 2FGL was not the final version of \citet{LAT11_J1825}.
  \item \citet{HB21_Ambrogi19} have shown that the previous template for the SNR HB 21 was too broad, because it mistakenly included a separate point source. We adopted their smaller template.
  \item \citet{VERJ2227_Xin19} reported the detection with \Fermilat of the extended VERITAS source VER J2227+608, corresponding to the SNR G106.3+2.7 next to the much brighter pulsar PSR J2229+6114. We have included it in the DR3 model.
\end{itemize}

\startlongtable
\begin{deluxetable*}{llllcl}
\tabletypesize{\scriptsize}
\tablecaption{Extended Sources Entered or Modified in the 4FGL-DR3 Analysis
\label{tbl:extended}}
\tablewidth{0pt}
\tablehead{
\colhead{DR3 Name}&
\colhead{Extended Source}&
\colhead{Origin}&
\colhead{Spatial Form}&
\colhead{Extent [deg]}&
\colhead{Reference}
}

\startdata
J0425.6+5522e & SNR G150.3+4.5 & 3FHL & Gaussian & 1.36 & \citet{G150_Devin20} \\
J0834.3$-$4542e & Vela X & 2FGL & Map & 1.00 & \citet{VelaX_Tibaldo18} \\
J1000.0$-$5312e & SNR G279.0+1.1 & New & Ring & 0.30, 1.44 & \citet{G279_Araya20} \\
J1640.7$-$4631e & HESS J1640$-$465 & New & Gaussian & 0.11 & \citet{HESSJ1640_Mares21} \\
J1824.4$-$1350e & HESS J1825$-$137 & 2FGL & Gaussian & 0.85 & \citet{LAT11_J1825} \\
J2044.9+5029e & HB 21 & 3FGL & Disk & 0.83 & \citet{HB21_Ambrogi19} \\
J2226.7+6052e & SNR G106.3+2.7 & New & Disk & 0.25 & \citet{VERJ2227_Xin19} \\
\enddata

\tablecomments{~List of updated and new sources that have been modeled as spatially extended. The Origin column gives the name of the \Fermilat catalog in which that extended source was first introduced (with a different template). The Extent column indicates the radius for Disk (flat disk) sources, the 68\% containment radius for Gaussian sources, the inner and outer radii for Ring (flat annulus) sources, and an approximate radius for Map (external template) sources.}

\end{deluxetable*}

Table~\ref{tbl:extended} lists the source name, origin, spatial template, and the reference for the dedicated analysis. These sources are tabulated with the point sources, with the only distinctions being that no position uncertainties are reported and their names end in \texttt{e} (see Appendix \ref{appendix_fits_format}). The new and modified extended sources have \texttt{DataRelease} set to 3. Unidentified point sources inside extended ones are indicated as ``xxx field'' in the \texttt{ASSOC2} column of the catalog, where ``xxx'' is the name of the extended source. A total of 78 extended sources are considered in DR3.

\subsection{Detection and Localization}
\label{catalog_detection}

\begin{figure}
\centering
\includegraphics[width=0.45\textwidth]{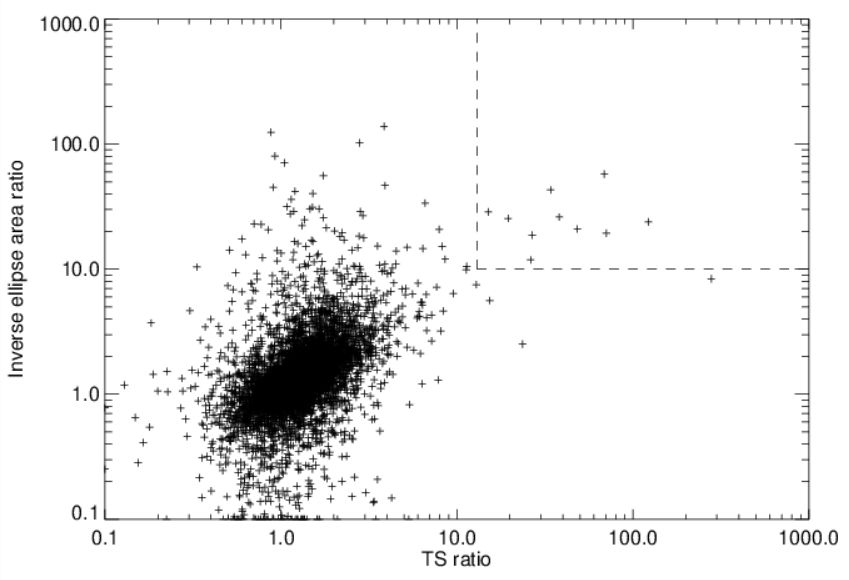}
\caption{Comparison of the $pointlike$ localizations of uw1216 seeds (for DR3) with those of uw8606 (for DR1). The abscissa is the ratio of TS values, whereas the ordinate is the inverse of the ratio of the areas of the error ellipses (such that larger is better). Sources in the upper right corner (materialized by the dashed lines) are stronger and much better localized in uw1216, and their positions and error ellipses were updated in DR3. The source at TS ratio around 300 is the bright $\gamma$-ray binary LS 5039, which appeared double in uw8606 because of its complex spectrum (resulting in low TS) but was well localized.}
\label{fig:ellipseratio}
\end{figure}

The source detection procedure followed the same lines as in 4FGL.
It used $pointlike$ \citep{Kerr2010} and a specialized diffuse model in which the nontemplate features are estimated differently.
It started from the previous list of sources (DR1 for DR2, DR2 for DR3), relocalized them over 12 yr of data, looked for peaks in the residual TS maps generated for several spectral shapes, introduced those in the model, refit, and iterated over the full procedure.
The resulting list (called uw1216) contained 11,128 seed sources at TS $>$ 10.

That procedure naturally resulted in changing the positions of all sources.
Since we wanted the catalog to be incremental, we forced most DR1 and DR2 sources back to their original positions (consistent with their names).
This requires associating the uw1216 seed sources with the DR2 catalog. To do that, we started by defining pairs in which each member is the nearest counterpart of the other, within at most 1$\degr$. We repeated that procedure after removing the pairs found at the first step, in order to handle faint sources close to much brighter ones (in which the nearest counterpart can be the bright source). We rejected 58 pairs in which the distance between the two members was larger than the largest of the two 99.9\% position errors (approximately 1.5 times the 95\% error) and $0\fdg15$. We eliminated the uw1216 seeds in a pair with a DR2 source, and formed the input list from the DR2 sources (including 139 with no counterpart in uw1216) and the remaining uw1216 seeds.

In DR2 we actually used the original positions of all DR1 sources strictly, and kept all of them.
However a number of sources became much brighter in recent years than they were during the first 8 yr, so they are now much better localized.
In recognition of this, we adopted for DR3 the new error ellipse when its area was at least ten times smaller and TS at least ten times larger than the same quantities in DR1. We used those very strict conditions to preserve the incremental nature of the catalog. This selected ten sources, clear outliers when looking at the distribution of ratios of ellipse areas and TS (Figure~\ref{fig:ellipseratio}). All are associated with blazars (very variable sources). One of them was in uw8606 (list of seeds for DR1) but too faint to be included in DR1 (it appeared in DR2), so only nine DR1 sources were updated.
They have a 4FGL counterpart in the \texttt{ASSOC\_FGL} column, but have \texttt{DataRelease} set to 3 because they have moved.

We have also adopted the new error ellipses of one DR1 source close to the modified extended source HESS J1825$-$137 and three faint soft sources (two from DR1, one from DR2) whose 95\% semi-major axes $R_{95}$ were larger than 1$\degr$ in DR1 or DR2.
Three DR1 sources with broad error ellipses $R_{95} \sim 0\fdg4$ were split into two better localized seeds inside the original error ellipse. We replaced the DR1 sources by the two seeds. The same applied to two DR2 sources.
Conversely we replaced two DR1 sources (4FGL J1831.5$-$0935 and J1830.2$-$1005) by a single uw1216 seed compatible with PSR J1831$-$0952.

We applied the same procedure as in 4FGL to eliminate seeds too close to a bright source and inside extended sources. Since we changed the extended sources (\S~\ref{catalog_extended}), this resulted in deleting several DR1 sources as well, not only new seeds. More precisely, 4FGL J1640.6$-$4632 was replaced by HESS J1640$-$465, five DR1 and two DR2 sources were deleted in or close to Vela X, three DR1 and one DR2 sources in the new SNR G279.0+1.1. We also deleted the faint 4FGL J1814.1$-$1710 next to the extended source HESS J1813$-$178 (even though this one did not change) because its TS decreased from 30 in DR1 to 16 in DR3, and it was not spectrally distinct from HESS J1813$-$178.

Following the $catXcheck$ verification (\S~\ref{catalog_catxcheck}) we have applied several modifications  to fit the data better. We replaced 4FGL J0857.7$-$4507 (in one of the clusters described in \S~\ref{lowlatunassocs}) by a stronger uw1216 source about $0\fdg5$ away.
We replaced 4FGL J1750.0$-$3849 just outside the extended source FHES J1741.6$-$3917 by a stronger point source just inside it.

We have looked at individual faint (TS $<$ 25) sources with a very soft ($\Gamma > 3.1$) spectrum, a large error ellipse ($R_{95} > 0\fdg5$) or a stronger close neighbor in uw1216 and deleted three DR1 (4FGL J0533.9+2838, J0313.6$-$7508, and J2326.5+8555) and three DR2 (4FGL J0750.0+7140, J1112.0+1021, and J1752.2$-$3002) sources. Nearby seeds appeared in DR3 and fit the data better. Similarly, we have deleted a faint very hard ($\Gamma < 1.3$) DR2 source (4FGL J1830.3$-$1601) that was superseded by a brighter nearby DR3 seed.

As before we manually added 11 known LAT pulsars that could not be localized by the automatic procedure without phase selection. Only one of those (PSR J1731$-$4744) reached TS $>$ 25 in DR3. It appears with a NULL error ellipse. The 10 others were discarded.
A total of 5029 new seeds were entered to the $gtlike$ source characterization for DR3 in addition to the 5690 remaining DR1 and DR2 point sources.

We reassessed the systematic corrections to localization. Up to 4FGL, they were defined globally over the entire sky, so their values were dominated by high-latitude sources (92\% of associated point sources in DR1 are at $|b| > 5\arcdeg$).
However, confusion and strong interstellar emission make localization much more difficult in the Galactic plane. In DR2 we estimated localization systematics separately at $|b| < 5\arcdeg$, based on 110 pulsars and low-latitude active galactic nuclei (AGN). We found that they should indeed be increased there, to $27\arcsec$ for the absolute 95\% error, and 1.37 for the systematic factor. We did not change the systematics in the high-latitude sky ($25\arcsec$ and 1.06). The DR2 systematics are used again for DR3.

\subsection{Thresholding and Light Curves}
\label{catalog_significance}

We used the Fermi Tools 1.4.7 analysis suite\footnote{See \url{https://fermi.gsfc.nasa.gov/ssc/data/analysis/documentation/}.}.
Compared to the Science Tools v11r7p0 that were used in 4FGL, the main difference is the new \texttt{edisp\_bins} setting for energy dispersion, which allows for broadening of the energy interval considered in the model space. The Science Tools setting was equivalent to \texttt{edisp\_bins} = $-$1, not enough to fully fill the first and last data bins for the small energy bins that we use (10 per decade). We used \texttt{edisp\_bins} = $-$2 instead, considering two bins for energy dispersion on each side beyond the data bins.
The acceptance for the PSF3 event type increases by a factor 2.4 between 50 and 100~MeV, so to account for this rapid change, we used 6 energy bins in the first component (Table~\ref{tab:components}) instead of 3 and set \texttt{edisp\_bins} to $-$4 accordingly.

\begin{deluxetable}{lrr}
\tablecaption{4FGL-DR3 priors on diffuse parameters
\label{tab:diffpriors}
}
\tablehead{
\colhead{Parameter Name} & \colhead{Mean} & \colhead{Std Dev} \\
}
\startdata
Galactic norm at 1~GeV, $K$ & 0.97 & 0.03  \\
Galactic spectral bias, $\Gamma_d$ & $-$0.01 & 0.02 \\
Isotropic normalization & 1.00 & 0.10 \\
\enddata
\tablecomments{The Galactic interstellar emission model is modulated by a power law $K (E/E_0)^{-\Gamma_d}$, in which $E_0$ is set to 1~GeV. $K$ tends to be smaller than 1. This is probably because of the sources that we added since the model was devised (just before 4FGL).
}
\end{deluxetable}

The likelihood weights were recomputed over 12 yr of data, resulting in slightly smaller weights everywhere.
We reoptimized all RoIs, resulting in 1988 RoIs containing up to 10 sources in their core. The maximum of 8 sources in the RoI core enforced in 4FGL was not optimal (it led to too many small neighboring RoIs and too many iterations).

For DR1 and DR2, the normalization of one of the diffuse components sometimes became very large (close to the maximum, set to 2) or very small (close to 0) in regions where that component contributed little. This was particularly obvious for the isotropic component in the Galactic ridge. This could happen because of missing features in the interstellar emission model (IEM), leading to a higher isotropic component in the fit. This was not satisfying since the isotropic emission is supposed to be approximately constant over the sky, and the stuck parameters caused trouble in the error estimates. In order to stabilize the diffuse parameters, we replaced the hard limits with Bayesian priors.
We chose the priors from the distributions of the parameters (in DR2) in regions where they are dominant ($|b| < 10\degr$ for the Galactic parameters and $|b| > 30\degr$ for the isotropic normalization). This led to the values reported in Table~\ref{tab:diffpriors}. The intrinsic reasons for the large or small values did not go away, of course, so the resulting isotropic normalizations were still larger than one in the Galactic ridge, but they rarely exceeded 1.5. The worst RoIs around $(l, b) \sim (315\degr, 3\degr)$ reached 1.7.

A major difference with the 4FGL procedure is that we retained all previous 4FGL sources (DR1 and DR2, except those discussed in \S~\ref{catalog_detection}), so that they would not be deleted from the model even if they have TS $<$ 25.
The resulting catalog contains 6659 entries, among which 1610 are new (1607 point sources plus the three extended sources discussed in \S~\ref{catalog_extended}), and 181 have TS $<$ 25 (112 DR1 and 69 DR2 sources). 

The light curves over 1 yr bins were recomputed (even the first 8 yr because adding new sources changes the fluxes of DR1 sources). The procedure was very similar to 4FGL, except we considered events up to 1 TeV, and we used a different model for the Sun and Moon at each year (see \S~\ref{DiffuseModel}). Neither change has a major effect.
Recomputing light curves over 2 month bins would have been very time consuming, and we showed in the 4FGL paper that the finer time binning was not critical to detecting variability. Monthly (and shorter intervals) light curves are provided in the \Fermilat Light Curve Repository.

The threshold on \texttt{Variability\_Index} is now 24.725 (same 1\% probability of false detection, applied to 12 bins).
The number of significantly variable sources increased from 1443 to 1695 (among which 158 are DR2 or DR3 sources). The fraction of variable sources decreased because part of the variability information in 4FGL came from the 2 month light curves. The distribution of fractional variability (\texttt{Frac\_Variability} in the FITS file) is similar to 4FGL, peaking between 50\% and 90\%. The blazars all have fractional variability larger than 10\% except 4C +55.17 (4FGL J0957.6+5523). The flux of that source, first discussed by \citet{LAT11_4C55}, decreased by about 13\% over 10 yr, not showing any flare.

\subsection{Spectral Fitting and Energy Distributions}
\label{catalog_spectra}

\begin{figure}
\centering
\includegraphics[width=0.45\textwidth]{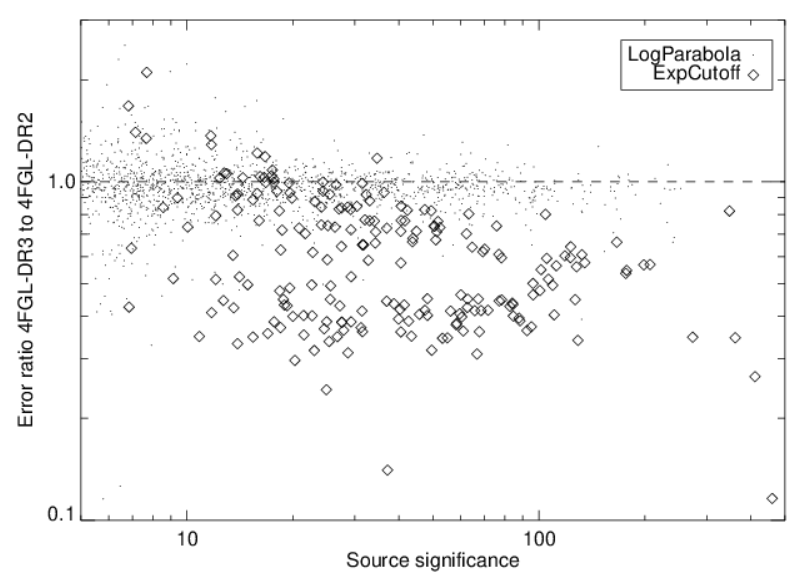}
\caption{Ratio of the error on the $\Gamma_S$ PLEC parameter (Eq.~\ref{eq:expcutoff}) in DR3 to that on the asymptotic low-energy index $\Gamma = \Gamma_S - d/b$ in DR2 (diamonds). The dots show the same error ratio on the $\alpha$ LP parameter (Eq.~\ref{eq:logparabola}), to gauge the effect of increased exposure (small). The large scatter at significance larger than 100 is due to the fact that those sources have free exponential index $b$ in DR3.}
\label{fig:plec_error}
\end{figure}

The spectral analysis proceeded along the same lines as in 4FGL, with five important differences detailed below: we changed the parameterization for pulsars, we fit more sources with four parameters, we changed the TS$_{\rm curv}$ threshold for using a curved spectral shape, we added new columns reporting the peak energy in $\nu F_\nu$, and we added one spectral bin to the spectral energy distributions (SEDs).

Most significantly curved sources are still fit with a lognormal function (\texttt{LogParabola} under \texttt{SpectrumType} in the FITS table, hereafter LP):
\begin{equation}
\frac{{\rm d}N}{{\rm d}E} = K \left (\frac{E}{E_0}\right )^{-\alpha -
\beta\ln(E/E_0)}.
\label{eq:logparabola}
\end{equation}

The significantly curved pulsars are still fit with a subexponentially cutoff power law (\texttt{PLSuperExpCutoff} under \texttt{SpectrumType} in the FITS table, hereafter PLEC). However we noticed that the parameters of the parameterization used in 4FGL (\texttt{PLSuperExpCutoff2} in the Fermi Tools) were strongly correlated, particularly when the exponential index ($b$ in Eq.~\ref{eq:expcutoff}) was free. This hampered convergence and precluded freeing $b$ except in the very brightest sources. To work around this, we devised a new parameterization (\texttt{PLSuperExpCutoff4} in the Fermi Tools):
\begin{eqnarray}
\frac{{\rm d}N}{{\rm d}E} & = & K \left (\frac{E}{E_0}\right )^{\frac{d}{b}-\Gamma_S} \exp \left [\frac{d}{b^2} \left (1 - \left (\frac{E}{E_0}\right )^b \right ) \right ]
\label{eq:expcutoff} \\
\frac{{\rm d}N}{{\rm d}E} & = & K \left (\frac{E}{E_0}\right )^{-\Gamma_S-\frac{d}{2}\ln\frac{E}{E_0}-\frac{db}{6}\ln^2\frac{E}{E_0}-\frac{db^2}{24}\ln^3\frac{E}{E_0}} {\rm for} \left | b \ln\frac{E}{E_0} \right | < 10^{-2},
\label{eq:expcutoff2}
\end{eqnarray}
in which the normalization $K$ is directly the flux density at the reference energy $E_0$ and the shape parameters are the spectral slope $\Gamma_S = {\rm d} \log (dN/dE) / {\rm d} \log E$ and the spectral curvature $d = {\rm d}^2 \log (dN/dE) / {\rm d} (\log E)^2$ at $E_0$.

The development in Eq.~\ref{eq:expcutoff2} shows explicitly that for $b$ = 0 the expression converges to LP with $d = 2\beta$. For $b < 0$ the shape (in log) is reversed (it decreases exponentially toward low energies and as a PL toward high energies). The parameters $K$, $\Gamma_S$, $d$ and $b$ appear as \texttt{PLEC\_Flux\_Density}, \texttt{PLEC\_IndexS}, \texttt{PLEC\_ExpfactorS} and \texttt{PLEC\_Exp\_Index} in the FITS table, respectively. The reference energy $E_0$ is chosen so that the error on $K$ is minimal and appears as \texttt{Pivot\_Energy} in the FITS table. This parameterization proved indeed much more stable, and the correlation between parameters is considerably reduced.

Figure~\ref{fig:plec_error} shows that on average $\Gamma_S$ is much better defined (error ratio below 1) than the low-energy asymptotic index $\Gamma = \Gamma_S - d/b$ used in 4FGL.
On average the error on $\alpha$ ($\Delta \alpha$) decreased by only 2.3\% between 10 and 12 yr, whereas $\Delta \Gamma_S$ is on average 0.626 $\Delta \Gamma$. Note that 15 pulsars had no $\Delta \Gamma$ at all in DR2 because $\Gamma$ was stuck at 0. This problem does not exist in the new parameterization. The small error ratio at significance below 40 corresponds to PSR J0622+3749. It had $\Gamma$ = 0.05 in DR2, very close to the boundary.
At low significance, a fraction of the pulsars have $\Delta \Gamma$ in DR2 that is similar or less than $\Delta \Gamma_S$ in DR3. The DR2 errors may have been underestimated because of the large correlation between parameters.

\begin{figure*}
   \centering
   \begin{tabular}{cc}
   \includegraphics[width=0.48\textwidth]{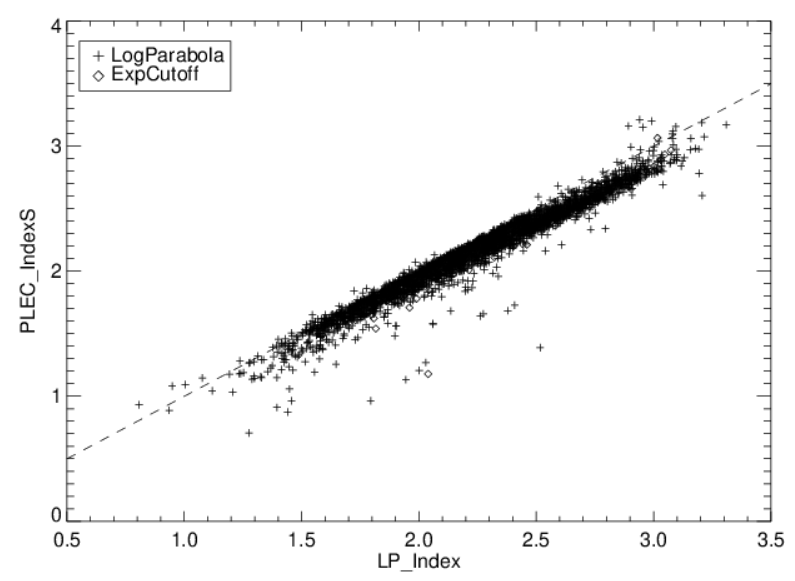} & 
   \includegraphics[width=0.48\textwidth]{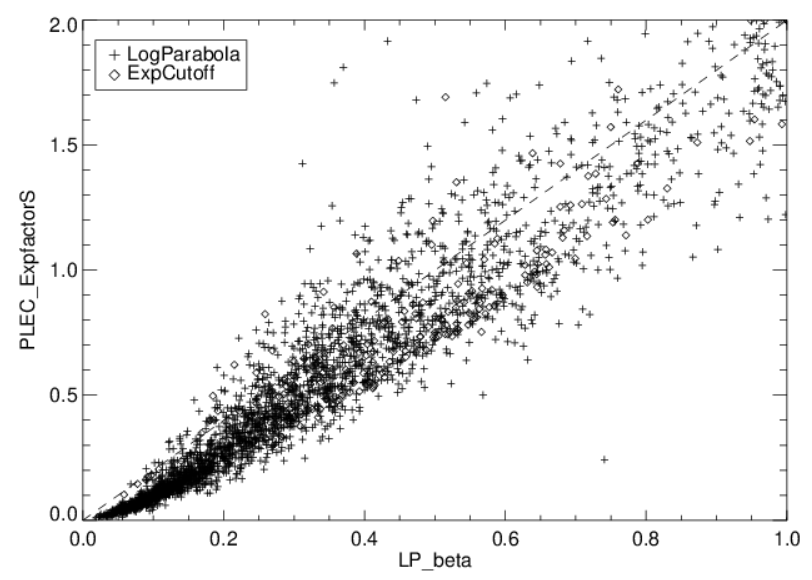}
   \end{tabular}
   \caption{Left: comparison of $\alpha$ of the LP model and $\Gamma_S$ of the PLEC model (\texttt{LP\_Index} and \texttt{PLEC\_IndexS} in the FITS file), showing that these parameters are largely similar.
     Right: comparison of LP $\beta$ and PLEC $d$ (\texttt{LP\_beta} and \texttt{PLEC\_ExpfactorS} in the FITS file), showing that $d$ is well correlated to $2 \beta$.
     The dashed lines show a one-to-one correlation.
 The outliers have large errors on both plots (none is farther off than 2 $\sigma$).}
\label{fig:Params_LP_PLEC}
\end{figure*}

Since the new PLEC parameters describe the same mathematical quantities (spectral slope and curvature) as the LP parameters (except the curvature is $2 \beta$), it is interesting to compare them. Both spectral models are fit to all sources, so this information is readily available.
Figure~\ref{fig:Params_LP_PLEC} shows that, as expected, the correlation between $\alpha$ and $\Gamma_S$ is quite tight.
The correlation between $2 \beta$ and $d$ is not as tight, because the curvature has larger error. On average $d < 2 \beta$, because the curvature in the PLEC model depends on energy, and the pivot energy (where $d$ is defined) is below the energy at which the two curvatures would be the same. This effect is stronger for bright sources that have lower pivot energy for the same spectral shape.

The new PLEC parameterization allows a 4-parameter fit (with free $b$) in fainter sources. We have applied it to all pulsars with TS $>$ 10,000 (28, up from 6 in DR1 and DR2). At TS $<$ 10,000 $\Delta b$ becomes larger than 0.15, which is the natural scatter on $b$ in the brightest pulsars, so freeing $b$ is no longer beneficial. For all other significantly curved pulsars, $b$ is fixed to 2/3 as in 4FGL. For comparison, the median $b$ over the 28 pulsars with free $b$ is 0.51, its weighted (with 1/$\sigma^2$) average is 0.55, and its intrinsic dispersion is 0.16. The Small Magellanic Cloud has $b$ = 1, as before \citep{LAT10_SMC}.

Besides 3C 454.3, the five other AGN with TS $>$ 80,000 were modeled with PLEC and free $b$ as well: CTA 102, Mkn 421, S5 0716+71, 3C 279 and PKS 1424$-$41. The new parameterization contains the LP model (for $b = 0$) so there is no risk of a worse fit, but nonconvergence can occur when the model is not constrained well enough. The TS threshold for free $b$ is higher in AGN than pulsars, because the curvature is much less in AGN, so $\Delta b$ is larger. The main objective of fitting PLEC with free $b$ remains to improve the modeling of the surroundings of very bright sources at low energy.

\begin{figure*}
   \centering
   \begin{tabular}{cc}
   \includegraphics[width=0.48\textwidth]{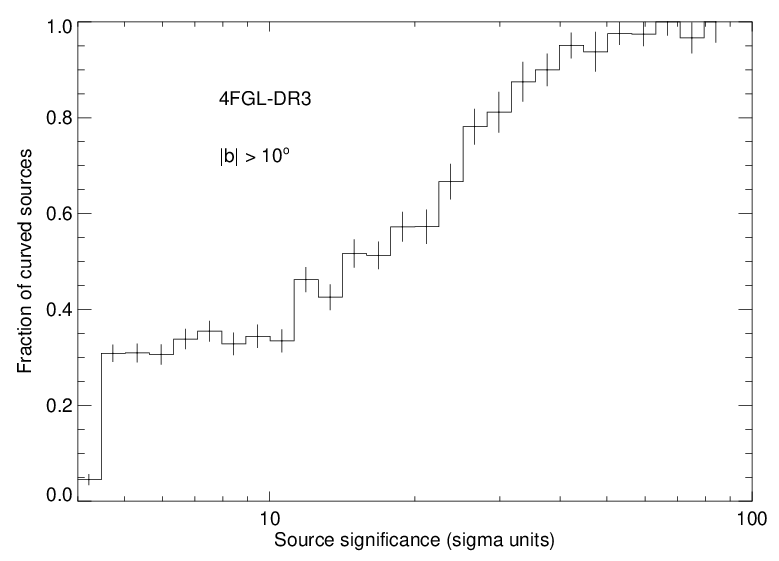} & 
   \includegraphics[width=0.48\textwidth]{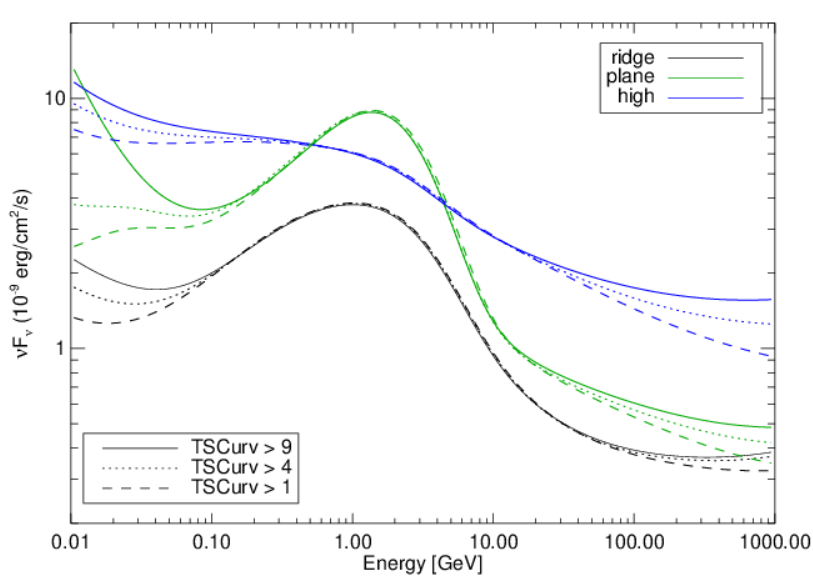}
   \end{tabular}
   \caption{Left: fraction of spectrally curved sources as a function of significance, outside the Galactic plane. The error bars are based on the source counts in each bin, assuming a binomial distribution. The sharp increase followed by a plateau at low significance (instead of the expected regular increase) is due to the fact that the source significance is boosted (moved to the right) for curved sources, combined with the detection threshold and a decreasing $dN/d \log TS$.
     Right: impact of changing the TS$_{\rm curv}$ threshold from 1 to 9 on the global source spectrum (sum of all sources, excluding the diffuse model) in three regions of the sky; high ($|b| > 10\degr$) in blue, ridge ($|b| \le 2\degr$ and $|l| < 60\degr$) in black, and plane ($|b| \le 10\degr$ excluding the ridge) in green. This test was carried out on the 10 yr data set and DR2 sources. The plot starts at 10~MeV, lower than the data (50~MeV).}
\label{fig:curved_frac}
\end{figure*}

In the DR1 and DR2 catalogs, the sources were represented with a curved spectral model (LP or PLEC) rather than a power law (PL) when TS$_{\rm curv} = 2 \, (\log \mathcal{L}$(curved spectrum)$ - \log \mathcal{L}$(PL)) was larger than 9 (3 $\sigma$). For DR3 we have lowered that threshold to 4 (2 $\sigma$) for two reasons, illustrated by Figure~\ref{fig:curved_frac}:
\begin{enumerate}
\item The left-hand plot shows that all sources are curved to some degree, when bright enough that the curvature is significant. This is physically understandable (the PL is an idealized asymptotic case).
\item The PL approximation tends to overestimate the source flux outside of the energy range in which the data are well constrained. The right-hand plot quantifies that on the sum of all sources, at three different TS$_{\rm curv}$ thresholds. Between 200~MeV and 20~GeV, it makes very little difference. High-latitude blazars (blue) are not strongly curved compared to pulsars and other Galactic sources (green and black). But above 20~GeV and below 200~MeV, the TS$_{\rm curv}$ threshold makes a difference. Most conspicuously, the global spectrum in the plane outside the ridge (green) diverges toward low energies at a threshold of 9 (what was actually used in DR2). The global spectrum is better behaved when the TS$_{\rm curv}$ threshold is lower.
\end{enumerate}
The implication is that we should choose a TS$_{\rm curv}$ threshold as low as possible while preserving global convergence of the model. For DR3 we set the threshold to 4 as a compromise because the test with a threshold at 1 required manual intervention, and global convergence could not be guaranteed. This is still a significant improvement as demonstrated in Figure~\ref{fig:curved_frac} (right).

Thanks to the improved statistics and the lowered TS$_{\rm curv}$ threshold, many more sources are considered significantly curved in DR3 than in DR1. The number of LP spectral models is now 3130 (among which 650 are new sources), while the number of PLEC models is now 258 (among which 2 are new sources). The fraction of sources represented by a curved spectral model increased from 30\% to 51\% overall.

\begin{figure}
\centering
\includegraphics[width=0.45\textwidth]{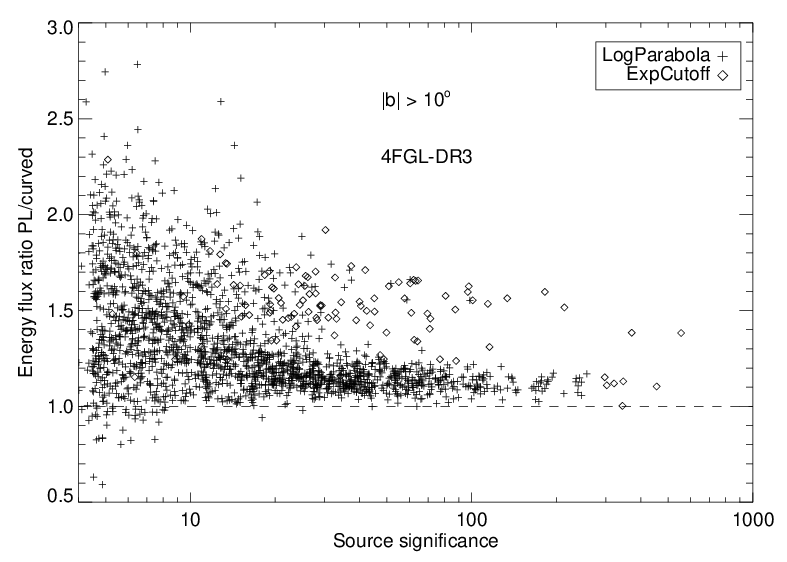}
\caption{Ratio between the energy flux (0.1 to 100 GeV) in the PL approximation and that in the LP or PLEC representation, for those sources that are considered significantly curved, outside the Galactic plane. Plus signs refer to the LP model, diamonds to the PLEC model. The scatter increases toward low significance because of the larger statistical fluctuations.}
\label{fig:Eflux_LP_PLEC}
\end{figure}

Changing the spectral models for many sources from PL to curved also has an impact on their energy flux estimate. The PL flux is nearly always larger, due to the integration down to 100~MeV for soft sources (or up to 100~GeV for hard sources $\Gamma < 2$). This is quantified in Figure~\ref{fig:Eflux_LP_PLEC}. Well-measured pulsars (diamonds at high significance) have on average a 50\% larger flux estimate in the PL model than in the PLEC model. The well-measured blazars (plus signs) are in the lower branch at a ratio around 1.1, because their spectra are much less curved than those of pulsars. The diamonds in the lower branch are the very bright blazars fit with PLEC (see above).
A ratio less than one occurs when the peak of the curved spectrum is close to either 100~MeV or 100~GeV (the boundaries of the energy interval over which the energy flux is extracted), because the full analysis interval (50 MeV to 1 TeV) extends beyond those boundaries and maintains the PL fit below the peak of the curved model.

To provide more information on curved spectra, we now report systematically in the catalog the peak energy in $\nu F_\nu$ and its uncertainty for all sources (including those not significantly curved) and both models as \texttt{(Unc\_)LP\_EPeak} and \texttt{(Unc\_)PLEC\_EPeak}:
\begin{eqnarray}
  E_{\rm peak}({\rm LP}) & = & E_0 \; \exp \left( \frac{2 - \alpha}{2 \, \beta} \right ) \\
  E_{\rm peak}({\rm PLEC}) & = & E_0 \; \left(1 + \frac{b}{d} (2 - \Gamma_S) \right )^{1/b}
\end{eqnarray}
The uncertainties are obtained using the covariance matrix.
We set those columns to NULL when the peak energy is undefined (when $\Gamma \ge 2$ in PLEC or curvature is upwards) or found too far from the LAT energy range ($|\ln (E_{\rm peak}/E_0)| > 10$). $E_{\rm peak}$(PLEC) is undefined in only 4 PLEC spectra, but in 516 LP and 1682 PL spectra (mostly because $\Gamma \ge 2$). $E_{\rm peak}$(LP) is always defined in PLEC and LP spectra, but remains undefined in 850 PL spectra in which $\beta$ is found negative or close to 0. The uncertainty on $E_{\rm peak}$(LP) is also set to NULL when $\beta$ is stuck to 1 and has no error (166 occurrences in LP spectra, 6 in PLEC spectra, and 3 in PL spectra).
Even when $E_{\rm peak}$ is formally defined, it should be treated with caution when $\Gamma$(PLEC) gets close to 2 or the curvature is close to 0.

\begin{figure}
\centering
\includegraphics[width=0.45\textwidth]{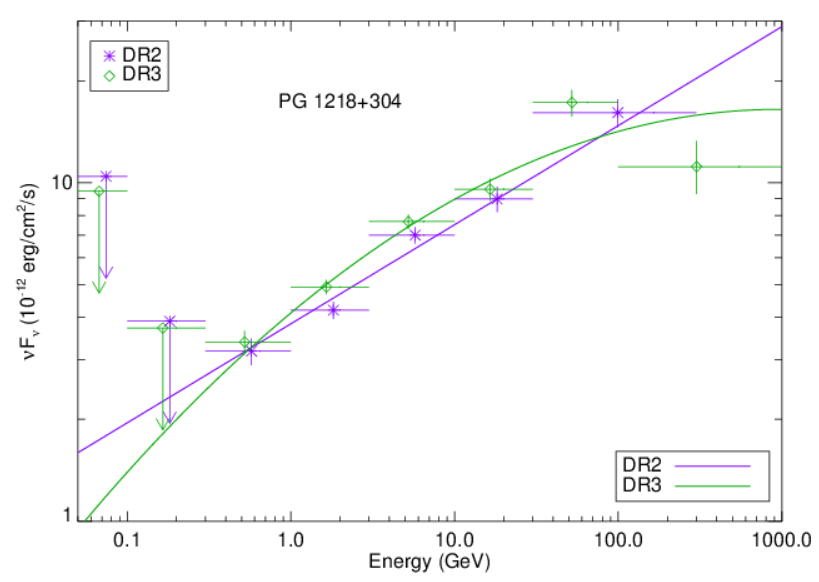}
\caption{SED of the BL Lac object PG 1218+304 (4FGL J1221.3+3010) for DR2 (purple crosses and PL fit) and DR3 (green diamonds and LP fit).}
\label{fig:SED_PG1218}
\end{figure}

In the DR1 and DR2 catalogs, we provided SEDs over $N_{\rm bands}$ = 7 energy bands. Thanks to the increased statistics after 12 yr (particularly valuable at high energy where the precision is limited by counts rather than background), we can provide them over 8 energy bands in DR3, extending to 1~TeV as in the 3FHL catalog \citep{LAT17_3FHL}. We replaced the 30 -- 300~GeV bin by a 30 -- 100~GeV bin and a 100~GeV -- 1~TeV bin.
We computed those two fluxes using unbinned likelihood, as in 4FGL. In the highest energy band, 172 sources have TS $>$ 25. An example of such a hard source is shown in Figure~\ref{fig:SED_PG1218}. That source has TS$_{\rm curv}$(LP) = 8.70, below the DR2 threshold of 9 but above the DR3 threshold of 4, so it is fit with the LP model in DR3. The significant differences between 1 and 10~GeV are due to the natural variability of that strong source.

\begin{figure}
\centering
\includegraphics[width=0.45\textwidth]{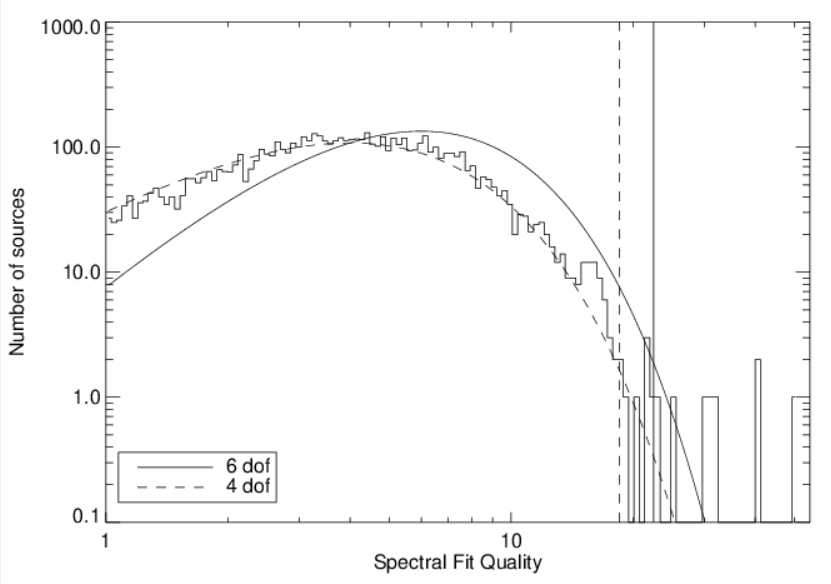}
\caption{Distribution of spectral fit quality over all DR3 sources (histogram). The $\chi^2$ distributions for 4 (dashed line, good fit) and 6 (solid line, bad fit) degrees of freedom are overlaid. The vertical lines correspond to a $p$-value of $10^{-3}$. The threshold for Flag 10 in Table~\ref{tab:flags} corresponds to the dashed vertical line (for 4 dof).}
\label{fig:SFQ_dof}
\end{figure}

The systematic uncertainties\footnote{See \url{https://fermi.gsfc.nasa.gov/ssc/data/analysis/LAT_caveats.html}.} have improved when moving from P8R3\_V2 to P8R3\_V3 and are now 0.1 in bands 1 and 8 and 0.05 in bands 2--7.
We compute the spectral fit quality (SFQ) as before \citep[Eq.~5 in][]{LAT20_4FGL}. Figure~\ref{fig:SFQ_dof} shows how that quantity is distributed. Up to DR2, we flagged (Flag 10 in Table~\ref{tab:flags}) sources with SFQ above the 99.9\% quantile ($10^{-3}$ $p$-value) of the $\chi^2$ distribution with $N_{\rm bands} - 2$ degrees of freedom (dof), considering that most sources were fit with PL spectra that have two free parameters. For DR3 that would result in 6 dof. Clearly the data do not follow the $\chi^2$(6) distribution, but are much closer to the $\chi^2$(4) distribution. This is due to two effects: first, half the sources are now fit with curved spectra (3 free parameters); second, the first ($<$ 100~MeV) and last ($>$ 100~GeV) bands do not behave as a Gaussian distribution (very large errors in the first band, and very low counts in the last band), so they contribute little to SFQ. For DR3 we flagged sources with SFQ $>$ 18.47, the $10^{-3}$ $p$-value of the $\chi^2$(4) distribution. The number of sources that trigger this remains small (only 16). The largest SFQ is obtained in PSR J0205+6449 at 49.6. That pulsar indeed has a known PWN (3C 58), which is very significant above the PLEC fit in the two bands above 30~GeV. The other sources flagged with Flag 10 have one spectral band off by more than 3 $\sigma$.

\subsection{Analysis Flags}
\label{catalog_analysis_flags}

\begin{deluxetable*}{crrrl}

\tablecaption{Comparison of the numbers of flagged sources between DR1 and DR3, separately for the DR1 sources (``DR1 in DR3'' column) and the sources introduced in DR2 and DR3 (``Post DR1'' column).
\label{tab:flags}}
\tablehead{
\colhead{Flag\tablenotemark{a}} & \colhead{DR1} & \colhead{DR1 in DR3} & \colhead{Post DR1} & \colhead{Meaning}
}

\startdata
  1  & 215 & 151 & 117 & $TS < 25$ with other model or analysis \\
  2  & 215 & 393 &  50 & Moved beyond 95\% error ellipse \\
  3  & 342 & 352 & 139 & Flux changed with other model or analysis \\
  4  & 212 & 233 & 227 & Source/background ratio $<$ 10\% \\
  5  & 398 & 418 & 259 & Confused \\
  6  &  92 & 161 & 156 & Interstellar gas clump (c sources) \\
  9  & 136 &  97 &  71 & Localization flag from {\it pointlike} \\
 10  &  27 &  44 &   2 & Bad spectral fit quality \\
 12  & 103 &  94 &  90 & Highly curved spectrum \\
 13  & \nodata & 112 & 69 & $TS < 25$ at 12 yr \\
 14  & \nodata & 368 & 181 & Soft Galactic Unassociated (\S~\ref{lowlatunassocs}) \\
 All & 1163 & 1433 & 720 & Any flag (\texttt{Flags} $>$ 0) \\
\enddata
 
\tablenotetext{a}{In the FITS file, the values are encoded as individual bits in the \texttt{Flags} column, with Flag $n$ having value $2^{(n-1)}$.}

\end{deluxetable*}

As before, we attempt to flag the sources affected by systematic errors larger than the statistical errors that we report, and which should therefore be treated with more caution. This can impact all source parameters, and even call into question the existence of the faintest flagged sources with significance less than 6$\sigma$ or 7$\sigma$ (TS $<$ 50). Because the largest source of systematics is the underlying diffuse emission, sources close to the Galactic plane, where diffuse emission is both the strongest and most structured, are most affected.

The flags (\texttt{Flags} in the FITS file) are recalled in Table~\ref{tab:flags} (see the 4FGL paper for the detailed definitions), together with the numbers of sources flagged for each reason and their evolution since DR1.
The effect of the underlying IEM was estimated by launching the procedure described in \S~\ref{catalog_significance} for a second time using the same seeds but the previous IEM (gll\_iem\_v06).

We note six changes with respect to the 4FGL procedure:
\begin{itemize}
\item For DR1, Flag 1 was applied to the comparison with both $pointlike$ and the alternative IEM. For DR3, the comparison to $pointlike$ was performed with no TS selection in order to apply Flag 3 to sources at TS $<$ 25. Flag 1 was determined only by the comparison with the old IEM, after selecting TS $>$ 25 on both sides. 
\item Flag 2 mostly relies on the comparison between the new uw1216 $pointlike$ positions of the 4FGL sources (\S~\ref{catalog_detection}) and the 4FGL positions (uw8606), because no localization was performed over 12 yr with the old IEM. The comparison with the old IEM also flags a few sources (although the seed positions are exactly the same) when among two nearby seeds one survives in DR3 and the other one with the old IEM.
\item All sources have been visually screened for potential deficiencies in the underlying diffuse model. The visual screening for diffuse features (Flag 6) depends on the source parameters. It was entirely redone for DR2. To be conservative, we left that flag (and the \texttt{c} suffix) in all the sources that had it in DR1, even if they would not have been flagged in DR2, and added the flag to all the new cases, including the sources already in DR1. For DR3 we checked only the new sources and did not change Flag 6 in the DR1 and DR2 sources. Two sources in the Large Magellanic Cloud (4FGL J0517.9$-$6930c and J0535.7$-$6604c) have that flag set as well.
\item Flag 12 still applies to PLEC sources with $\Gamma = \Gamma_S - d/b \le 0$ (as well as to LP sources with $\beta$ fixed to 1, as before), even though the new PLEC parameterization (Eq.~\ref{eq:expcutoff}) has no explicit limit at $\Gamma = 0$.
\item We introduced for DR2 a new Flag (13) to flag explicitly sources from a previous version with TS $<$ 25 in the current version. It now applies to DR2 sources as well (69 have TS $<$ 25 in DR3).
\item For DR3, we introduce a new Flag (14) to flag explicitly soft Galactic unassociated (SGU) sources in regions with high source density (\S~\ref{lowlatunassocs}). The three strongest SGUs (4FGL J1702.7$-$5655 at $b = -9.2$, 4FGL J1823.3$-$1340 inside HESS J1825$-$137, and 4FGL J1839.4$-$0553 inside HESS J1841$-$055), all at TS $>$ 800, are very clear point sources (good pulsar candidates) and are not flagged.
\end{itemize}
Overall the fraction of sources with any flag set increased from 23\% in DR1 to 32\% in DR3. Among the new DR2 and DR3 sources, 75\% are flagged in the Galactic plane ($|b| < 10\arcdeg$), but only 26\% are flagged above the Galactic plane.

\subsection{All-sky verification}
\label{catalog_catxcheck}

In order to verify that the 4FGL-DR3 catalog provides a good representation of the $\gamma$-ray sky, we have performed an independent all-sky analysis, named $catXcheck$, using the $p$-value statistic (PS) data/model deviation estimator developed by~\citet{PSmap}. $catXcheck$ consists of the analysis of 540 RoIs ($12\times12\degr$, with a pixel size of $0\fdg1$) covering the whole sky. The RoI centers lie on 19 Galactic parallels, whose latitudes range from $-90\degr$ to $90\degr$ with a $9\degr$ step. The longitude step is $9\degr$ for $|b| \leq 18\degr$, $10\degr$ for $|b|=27\degr$ and $36\degr$, and it is 12$\degr$, 15$\degr$, 18$\degr$, 24$\degr$, and $45\degr$ for $|b|=45\degr$, 54$\degr$, 63$\degr$, 72$\degr$, and $81\degr$, respectively. The difference with respect to the catalog main analysis is that we combine all PSF event types together, using all events whose zenith angle is less than $90\degr$ to suppress contamination from Earth limb $\gamma$ rays. The energy range is [100 MeV, 1 TeV], and we use 10~bins per decade.

The source model of each RoI comprises the Galactic diffuse emission, the isotropic template, the Sun and Moon steady emission templates, as well as all point-like and extended sources from DR3, within $5+0.015(\sigma_\mathrm{src}-4)$~degrees of the RoI border, where $\sigma_\mathrm{src}$ is the source significance (\texttt{Signif\_Avg} in the FITS version of the catalog). That formula selects sources outside of the RoI up to 5$\degr$ for the faintest ones and to 15$\degr$ for the brightest ones, corresponding respectively to the 68\% and 95\% containment radii at 100~MeV.

Since we want to check the goodness of the DR3 information, the spectral parameters of the DR3 sources are fixed to their catalog values, and the only free parameters of the binned-likelihood fit are the ones of the Galactic diffuse emission (power-law correction) and isotropic (normalization) templates. After the fit, we compute the 3D map of the number of photons predicted by the source model and compute the PS map in order to detect any significant deviation between the observed and predicted 3D maps.

As explained in~\citet{PSmap}, the PS estimator is computed for each spatial pixel of a RoI, and rather than comparing the observed and predicted count spectra at that pixel, we compare spatially integrated count spectra around that pixel, with an integration radius varying with energy as $p_0 (E/100~\mathrm{MeV})^{-p_1} \oplus p_2$, where $\oplus$ is the sum in quadrature. We use the point-source optimized integration parameterization (which is very close to the PSF 68\% containment energy dependence): $p_0=4\degr$, $p_1 = 0.9$ and $p_2=0\fdg1$, and we merge the energy bins to have an energy binning of 0.3 in $\log_{10}{E}$. PS is defined such that $\mathrm{|PS|} = -\log_{10}(p \text{-value})$, where the $p$-value is the probability that the statistical fluctuations can reach a level of deviation as large as the one observed in the data, under the assumption that the model represents the data. The PS sign is the sign of the sum of the count spectra residuals in units of $\sigma$. The likelihood weights are taken into account in the PS computation. We chose to use the PS map to investigate data/model deviations because, contrary to the TS map, it is sensitive to both positive and negative deviations.

\begin{figure}
\centering
\includegraphics[width=0.5\textwidth]{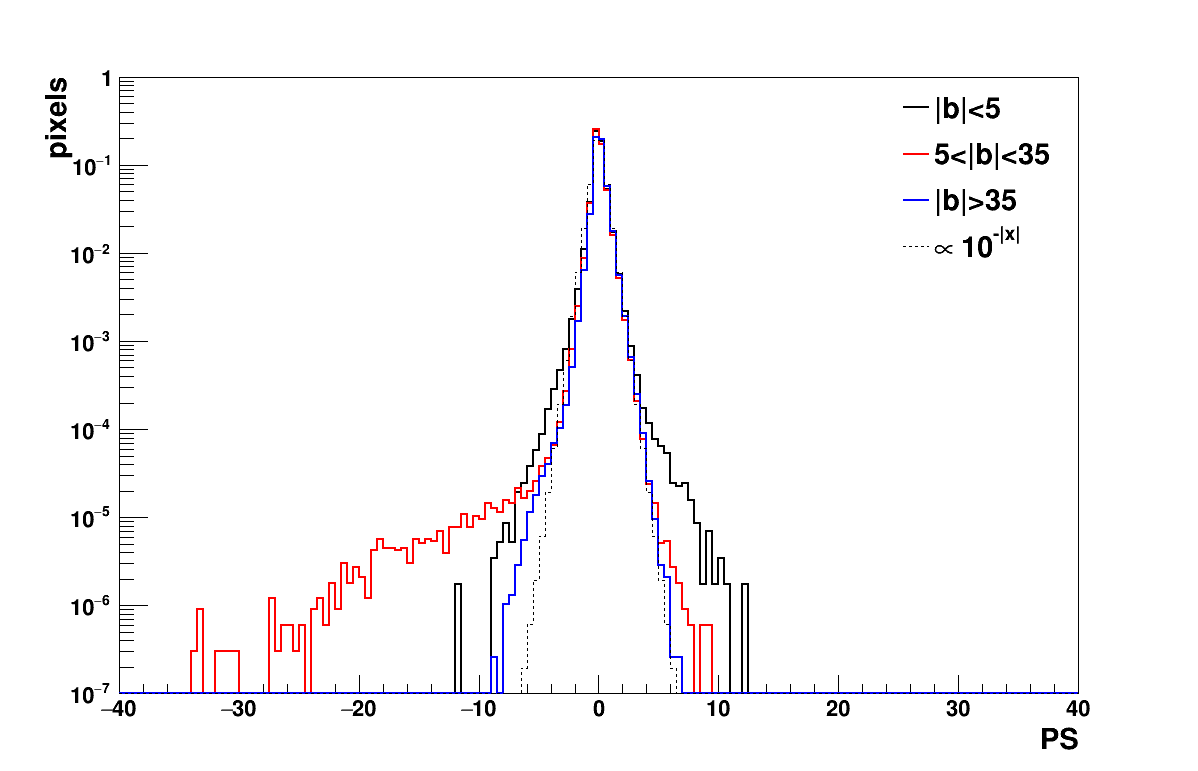}
\caption{PS distribution for the three Galactic-latitude selections. The thin solid histogram corresponds to the expected $10^{-|x|}$ distribution. All histograms are normalized such that their integral is 1.}
\label{fig:catXcheck_optsrc_PS1d}
\end{figure}

For one pixel, the $3\sigma$, $4\sigma$, and $5\sigma$ thresholds correspond to PS = 2.57, 4.20 and 6.24. Since the level of correlation between pixels is negligible with $p_2=0\fdg1$, we only have to take into account the number of trials when deriving the all-sky significance thresholds. The number of RoIs and their centers have been chosen so that the overlap between neighboring RoIs ensures that, for any given direction in the sky, there is at least one RoI containing this direction and providing a distance to the RoI border larger than $\sim 1\fdg5$. As a consequence, when combining the results of the 540 PS maps, we loop over the RoIs and the RoI pixels, and we only consider the pixels for which the RoI provides the largest distance to the RoI border. This allows us to avoid double counting the sky directions. The total number of considered pixels is 4,175,468 and adding its logarithm to the one-pixel significance thresholds gives the all-sky $3\sigma$, $4\sigma$, and $5\sigma$ thresholds: PS = 9.19, 10.82, and 12.86.

\begin{figure}
\centering
\includegraphics[width=0.5\textwidth]{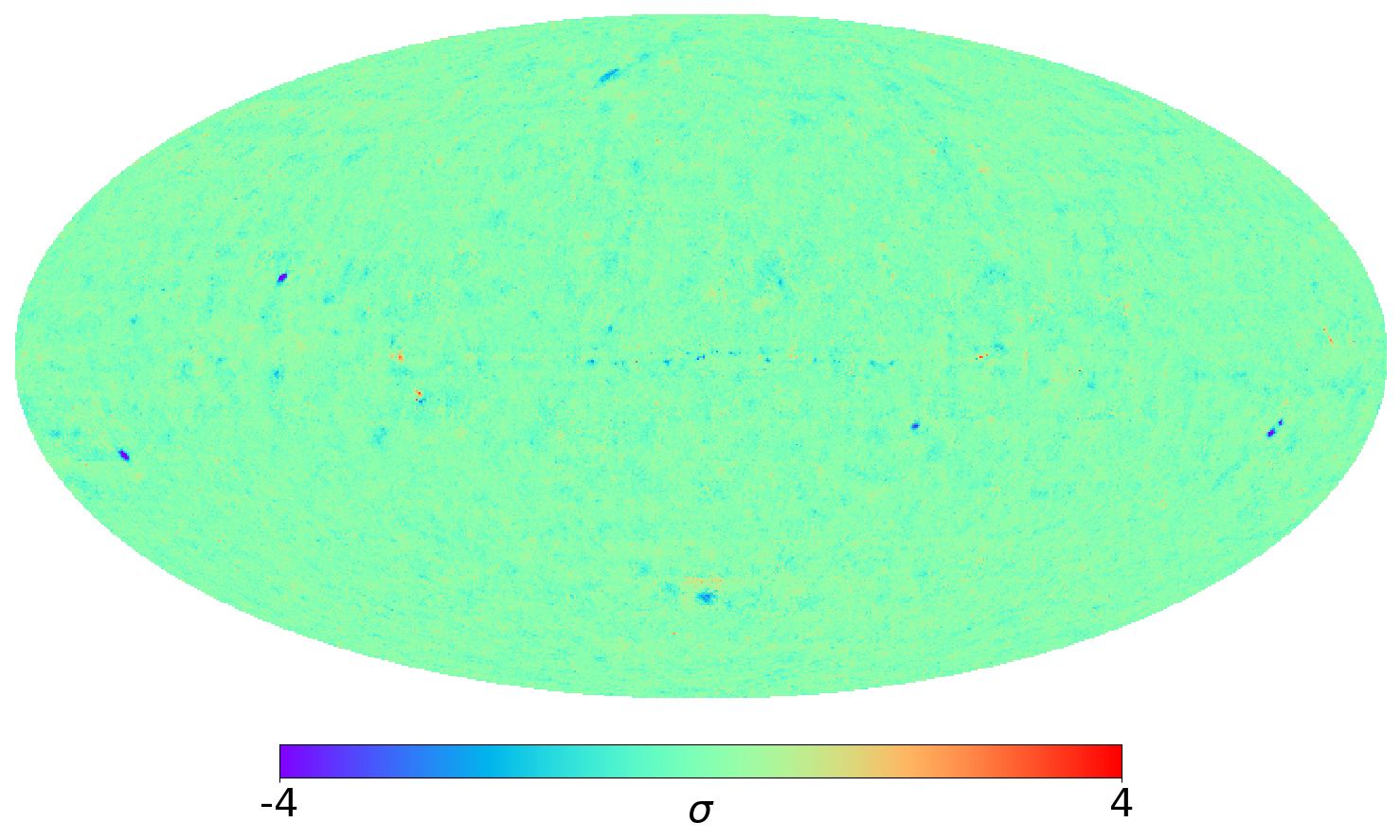}
\caption{All-sky PS map with `source' settings in Galactic coordinates (Mollweide projection). The color scale corresponds to $-4\sigma$ to $4\sigma$ deviations (accounting for the number of pixels in the map).}
\label{fig:catXcheck_optsrc_PSallsky}
\end{figure}

The PS distributions for samples of sources in three different ranges of Galactic latitude are shown in Figure~\ref{fig:catXcheck_optsrc_PS1d}. The high-latitude ($|b|>35\degr$) distribution follows the $10^{-|x|}$ expectation, whereas some deviations are visible in the low-latitude ($|b|<5\degr$) and mid-latitude ($5\degr<|b|<35\degr$) distributions. To show the locations of these deviations, we build an all-sky HEALPix~\citep{Gorski2005} map in Galactic coordinates with $N_\mathrm{side}=256$ from the 540 PS maps. Figure~\ref{fig:catXcheck_optsrc_PSallsky} shows the resulting all-sky PS map. Overall the data/model agreement is good over the whole sky. Using a $4\sigma$ threshold, we find two positive and five negative deviations. The positive ones are both located inside bright extended sources: 4FGL~J1923.2$+$1408e (W~51C) and 4FGL~J0822.1$-$4253e (Puppis~A). They occur because the geometric templates that we use for those sources are far from perfect, and we do not allow additional point sources inside them (\S~\ref{catalog_detection}). Regarding the negative deviations, the one at a distance of $\sim 1\degr$ from the Galactic center corresponds to only one pixel. The other four are clusters of several pixels with $\mathrm{PS}<-10.82$ and correspond to extended deviations explaining the negative tail in the mid-latitude histogram in Figure~\ref{fig:catXcheck_optsrc_PS1d}. They all are close to large molecular clouds: one can be associated with Cepheus, one with Perseus, and two with Orion~B.

\begin{figure}
\centering
\includegraphics[width=0.5\textwidth]{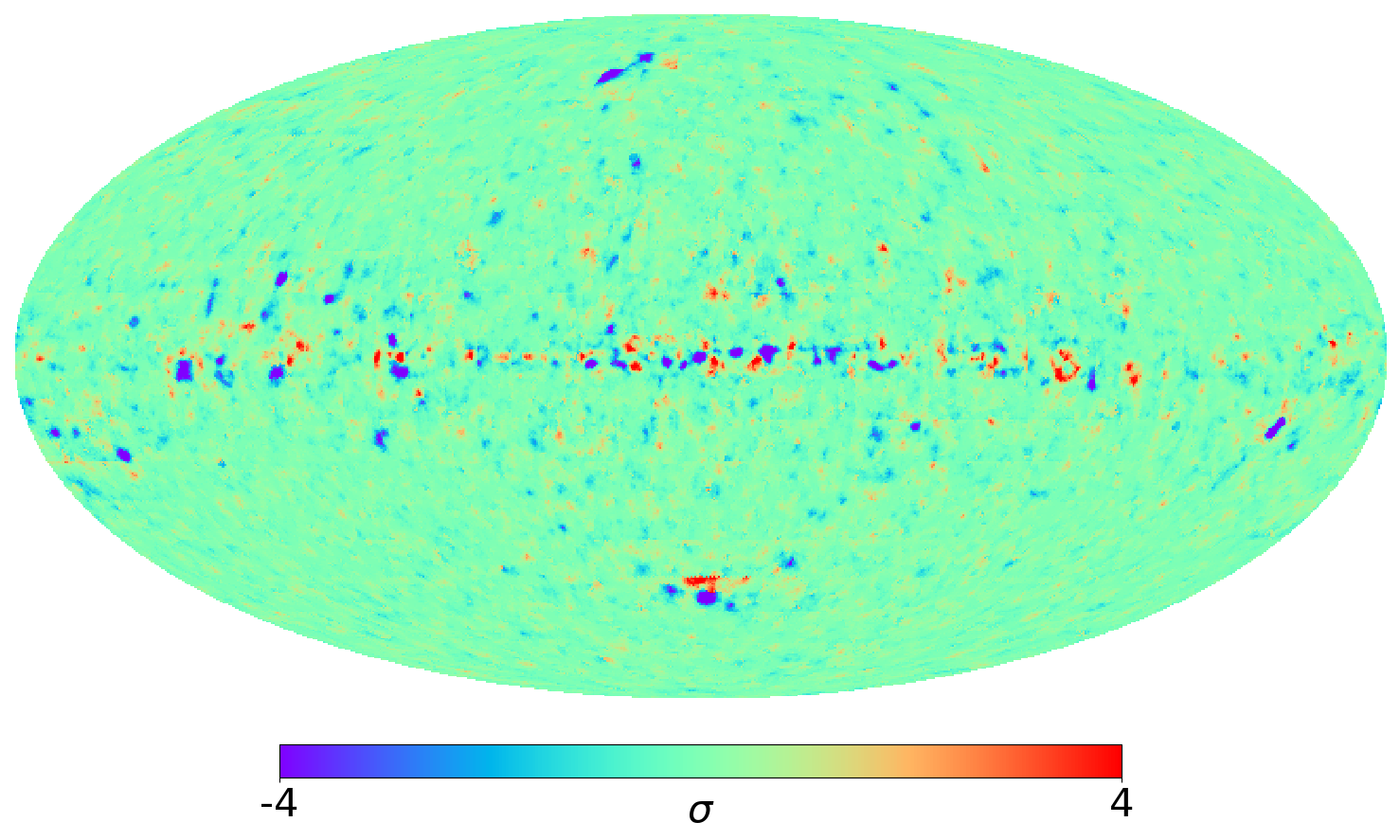}
\caption{All-sky PS map  with `diffuse' settings in Galactic coordinates (Mollweide projection). The color scale corresponds to $-4\sigma$ to $4\sigma$ deviations (accounting for the number of independent pixels in the map).}
\label{fig:catXcheck_optdif_PSallsky}
\end{figure}

Since these extended negative deviations seem to be related to the Galactic diffuse emission, we also compute the 540 PS maps with the set of PS spatial integration parameters optimized for large extended deviations ($p_0=5\degr$, $p_1 = 0.8$ and $p_2=1\degr$). With $p_2=1\degr$, the level of correlation between pixels is expected to increase. Assuming that pixels within $0\fdg5$ of each other are highly correlated, we divide the number of trials by 69, and the $3\sigma$, $4\sigma$, $5\sigma$ thresholds become PS = 7.4, 9, 11. The corresponding all-sky PS map, shown in Figure~\ref{fig:catXcheck_optdif_PSallsky}, exhibits many deviations above $4\sigma$. A preliminary analysis of the negative ones shows that many of them are correlated with some components of the IEM, especially the local CO template (Galactocentric radius range 7--9 kpc), which may indicate that allowance for variation of the X$_{\rm CO}$ factor among the local clouds is necessary. The positive deviations are less numerous and less significant. Like the negative ones, some are likely due to imperfections in the modeling of the Galactic diffuse emission, but some could correspond to new extended sources.

Although a complete analysis of these extended deviations is outside the scope of this paper, we try to quantify their impact on the DR3 catalog. In order to do so, we look at the 540 PS maps and count the number of pixels with $\mathrm{|PS|} > 7.4$ and derive the expected number of DR3 sources in these regions from the average source density in the $\mathrm{|PS|} < 7.4$ region. In the 40 RoIs centered along the Galactic plane, the fraction of negative and positive deviation pixels is 2\% and 0.9\%, respectively. The numbers of observed and expected DR3 sources are 15 and 24.2 in the negative deviation pixels, and 33 and 10.1 in the positive deviation pixels. In the RoIs farther from the Galactic plane, the fraction of negative and positive deviation pixels is 0.15\% and 0.03\%, respectively. There are no DR3 sources in  the negative deviation pixels, while 8.4 are expected. We conclude that the catalog results are biased in the negative and positive deviation regions, but the effect is small and the number of missed or spurious sources is at most $\sim 20$.

\section{The 4FGL-DR3 Catalog}
\label{dr3_description}

The catalog is available online\footnote{See \url{https://fermi.gsfc.nasa.gov/ssc/data/access/lat/12yr_catalog/}.}, together with associated products.
It contains 6659 entries (1607 new since DR1).
The source designation is \texttt{4FGL JHHMM.m+DDMM}.
The format is similar to 4FGL (except it does not contain the 2 month light curves) but not identical. It is detailed in App.~\ref{appendix_fits_format}.
A new column (\texttt{DataRelease}) is set to 1 for the unchanged DR1 sources, 2 for DR2 sources, and 3 for DR3 sources. Sources existing in DR1 or DR2 but whose position (and therefore name) changed (\S~\ref{catalog_detection}) have \texttt{DataRelease} set to 3. The DR1 or DR2 names of those sources are given in the \texttt{ASSOC\_4FGL} column of the FITS file. They are considered DR1 or DR2 sources in the text, tables, and plots.

\subsection{The 4FGL-DR1 sources}
\label{4fglsources}

\begin{figure}
\centering
\includegraphics[width=0.45\textwidth]{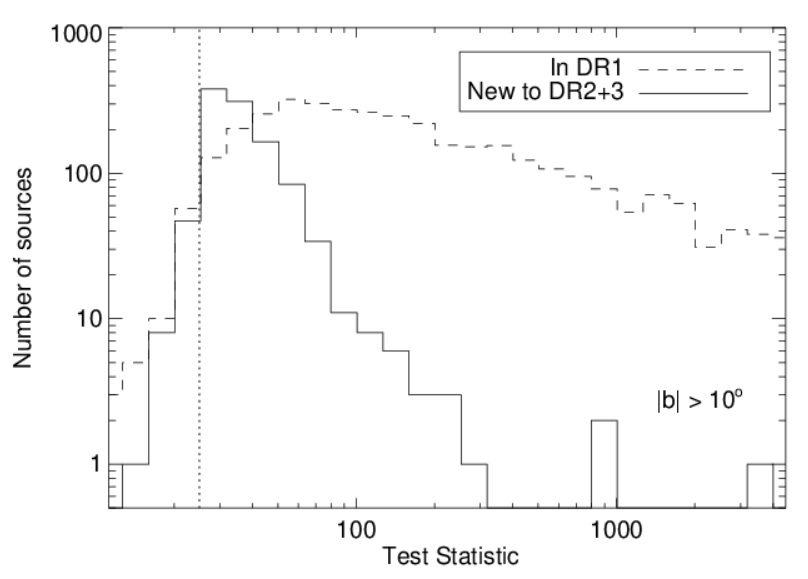}
\caption{Distributions of the 4FGL-DR3 Test Statistic at high latitude ($|b|>10\degr$) for the sources already in DR1 (dashed) and those new to DR2 or DR3 (solid). The vertical dotted line is the threshold at TS = 25. The solid histogram extends below the threshold because 69 DR2 sources have TS $<$ 25 in DR3, including 46 at $|b| > 10\degr$.}
\label{fig:TS_comp}
\end{figure}

The TS of the 5048 remaining DR1 sources in DR3 increased on average compared to DR1. Among sources that are fit with the PL model in both catalogs (no TS boost from switching to a curved model), TS increased on average by 33\% at high latitude ($|b| > 10\degr$) and by 22\% closer to the Galactic plane.
The difference between the two is due to the lower log-likelihood weights, which partly offset the exposure increase close to the Galactic plane where diffuse emission is strong, particularly for soft sources.
Even at high latitude, the TS ratio is less than expected from the exposure increase (50\%). This is not due to the lower weights (the TS ratio does not depend on index) but to the selection bias (sources tend to be brighter in the interval in which they were defined) and signal-splitting with new sources.
The statistical uncertainties on the parameters decreased by 15\% at high latitude and 9\% close to the plane.

The random character of adding new data and the variable nature of the $\gamma$-ray sources inevitably lead to broadening the TS distribution, in addition to shifting it to larger values. The sharp selection threshold at TS = 25 in DR1 becomes a tail of sources at TS $<$ 25 in DR3. Figure~\ref{fig:TS_comp} shows the TS distribution outside the Galactic plane, where 74 DR1 sources have TS $<$ 25.
Over the entire sky, 112 DR1 sources have TS $<$ 25 in DR3, among which 15 have TS $<$ 16, and the smallest TS is 9.8 (4FGL J1816.4$-$0057).

The energy flux decreased on average by 5\% (0.2 $\sigma$) between DR1 and DR3 (considering only PL sources to avoid the effect of switching to a curved model). This is again a selection bias dominated by faint sources: close to the detection threshold, DR1 sources can get either brighter or fainter in DR3, but faint sources in DR3 cannot be fainter in DR1 because they would have TS $<$ 25. Real variability is apparent for bright sources (TS $>$ 1000). The scatter on the energy flux ratio is 5 $\sigma$ on variable DR1 sources, but only 1 $\sigma$ on nonvariable ones. There is no offset and less scatter (only 0.8 $\sigma$ overall) on the difference of PL indices because source fluxes vary strongly, but spectral shapes only slightly.

\subsection{The new DR2 and DR3 sources}
\label{dr2sources}

\begin{figure*}[!ht]
   \centering
   \begin{tabular}{cc}
   \includegraphics[width=0.48\textwidth]{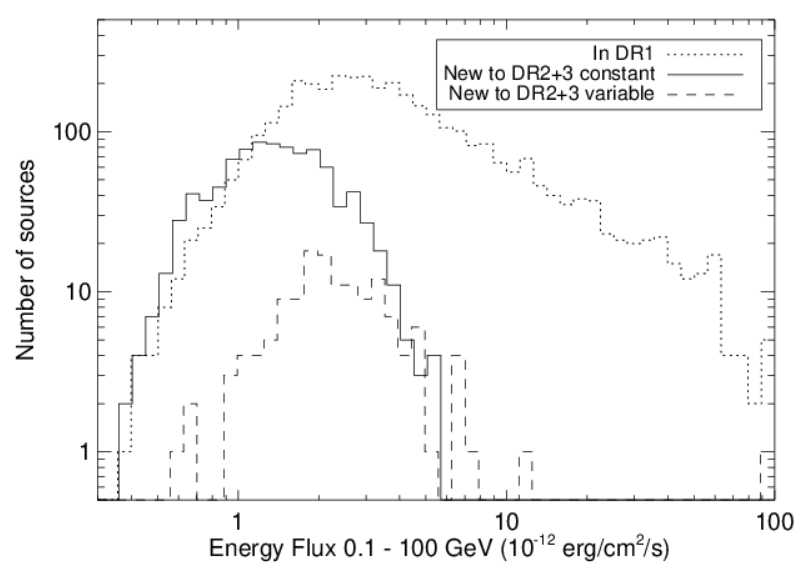} & 
   \includegraphics[width=0.48\textwidth]{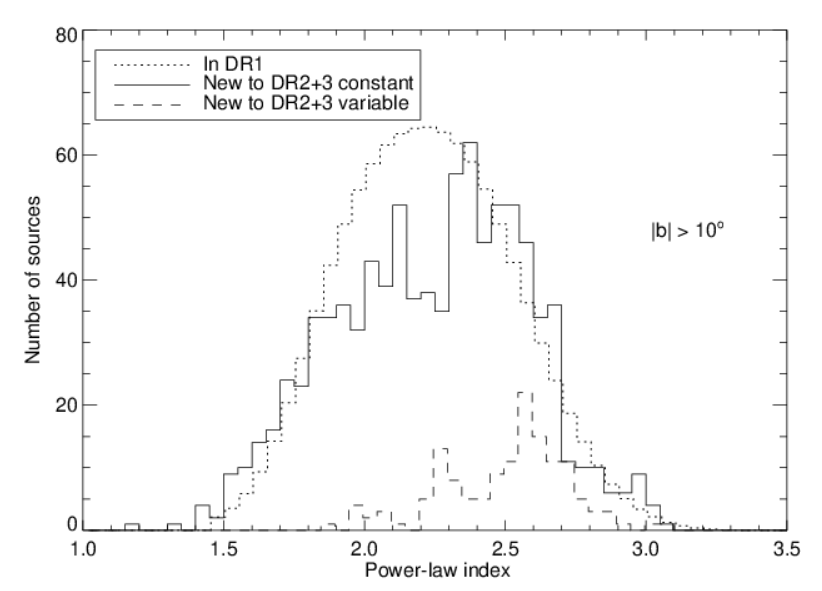}
   \end{tabular}
   \caption{Left: distributions of the 4FGL-DR3 energy flux at high latitude ($|b|>10\degr$) for the sources already in DR1 (dotted) and those new to DR2 or DR3, separately for those considered variable (dashed) or not (solid). The energy flux is obtained with the best spectral model, which can be either a curved model or a power law (\S~\ref{catalog_spectra}).
     Right: same distributions for the PL index (always obtained with the power-law model). The DR1 distribution was broadened to account for the larger uncertainties of the index values for the new sources, so the comparison is fair.}
\label{fig:flux_index_comp}
\end{figure*}

The 1607 new DR2 and DR3 sources are on average very close to the detection threshold (median TS of 34). This is obvious in Figure~\ref{fig:TS_comp}.
There are a few newly active bright blazars at TS $>$ 100, and even three at TS $>$ 800 (4FGL J0154.5$-$6604, J1544.3$-$0649, and J2250.0$-$1250 associated with PKS 0153$-$663, NVSS J154419$-$064913, and PKS 2247$-$131 respectively) but the majority are faint sources that rise above the detection threshold thanks to the additional four years.
Outside the Galactic plane, the confusion limit is still distant, because ever-deeper exposures reach to higher energies at which the PSF is better. Among the 1042 faint (TS $<$ 100) new sources at $|b| > 10\degr$, only 7\% have the confusion flag (Flag 5 in Table~\ref{tab:flags}) set.

Figure~\ref{fig:flux_index_comp} (left) shows that the energy flux distribution of the new sources peaks between 1 and $2 \times 10^{-12}$ erg cm$^{-2}$ s$^{-1}$.
The majority (140) of the 156 new DR2 and DR3 sources considered variable (mostly blazars) are outside the Galactic plane ($|b| > 10\arcdeg$). Their average flux is larger than that of the nonvariable sources. This is largely a selection bias (variability cannot be detected in the very faint sources).

Figure~\ref{fig:flux_index_comp} (right) compares the distribution of PL indices of the new sources with that of the DR1 ones.
The variable new sources are soft on average, similar to flat-spectrum radio quasars (FSRQ). The same effect is observed in the 1FLT catalog \citep{2021_1FLT}, even more strongly because detections in 1FLT are based on one-month data sets.
The nonvariable new sources have a peak on the soft side as well (probably similar to the variable ones, but not bright enough to be significantly variable).
But they also have a shoulder on the hard side. This is because adding more exposure is more beneficial at high energies (in the count-dominated regime) than at low energies, and therefore helps with detecting hard sources (like BL Lac objects).

\subsection{Step-by-step from DR1 to DR3}
\label{stepbystep}

To understand the effects of the analysis changes introduced in DR3 with respect to 4FGL, we have considered their effects on the DR2 data set (ten years). To that end, we started with the same seeds as the DR2 catalog, changed each element in sequence (in the order of the list below), and compared each intermediate result with the previous one.
\begin{itemize}
\item We first switched from the Science Tools to the Fermi Tools, and from the P8R3\_V2 to the P8R3\_V3 data and Instrument Response Functions (changing the isotropic templates accordingly). This led to essentially no change at all, as expected.
\item Setting \texttt{edisp\_bins} to $-$2 (\S~\ref{catalog_significance}) when accounting for energy dispersion had a significant effect. Prior to that, the model counts were underestimated by about 5\% in the first energy band (50 to 100~MeV), so the model fit was too high by the same amount at low energy. We found 16 fewer sources with the new settings, due to a slight TS reduction in soft sources. The energy flux decreased by 2\% on average. The largest effect was spectral: we found 155 more LP spectra (+ 10\%). Hard PL sources ($\Gamma < 2.1$) were unchanged, but $\Gamma$ decreased more and more for softer sources, reaching $\Delta \Gamma = -0.1$ at $\Gamma = 3$.  Note that this is specific to the catalog setting that uses many components for the summed likelihood in relatively narrow energy bands (Table~\ref{tab:components}). Analyses using few broader energy bands are not as sensitive to \texttt{edisp\_bins}.
\item Introducing diffuse priors had very little effect on the sources. We found 7 fewer sources, 11 fewer LP spectra. In the Galactic ridge (where the effect of the priors is strongest), the average energy flux of sources decreased by 2\% (0.1 $\sigma$), and the average PL index decreased (became harder) by 0.005 (0.06 $\sigma$). This is completely innocuous.
\item Lowering the TS$_{\rm curv}$ threshold from 9 to 4 (\S~\ref{catalog_spectra}) had an important effect on the sources, besides the collective effect illustrated in Figure~\ref{fig:curved_frac} (right). The number of LP spectra increased by 1000 (+ 58\%). The total number of sources increased by 127 (+ 2\%) because more LP sources allowed more soft neighbors to reach the detection threshold. The energy fluxes of sources that switched from PL to LP decreased by 1.8 $\sigma$ (well above the statistical precision) as illustrated in Figure~\ref{fig:Eflux_LP_PLEC}. Their neighbors picked up most of that flux, so their energy fluxes increased. They also became somewhat softer in the process (by 0.02 on average). On the other hand, the neighbors of those neighbors became a little harder, so that the global average PL index did not change.
\end{itemize}
In conclusion, the change with the greatest consequences was lowering the TS$_{\rm curv}$ threshold. We are confident that the new setting is closer to the physical reality.

\section{Associations}

\label{dr3_assocs}
We have performed the same association procedure as in 4FGL, using the Bayesian method and likelihood-ratio method \citep[see][]{LAT20_4FGL}. The procedure yielded associations for 341 new DR2 sources and 366 new DR3 sources, representing association fractions of 48\% and 41\% respectively. These fractions are roughly consistent with that expected from DR1 for these low-TS sources. We provide low-probability ($0.1<P<0.8$) associations for 95 new DR2 or DR3 sources and associations with Planck sources for 44 others. 
One major change regarding classes is that now we split the generic pulsar class used in previous catalogs into two separate classes, namely young pulsars (PSR) and millisecond pulsars (MSP). We recall that some pulsars with LAT-detected pulsations do not fill the defining significance criterion of this general source catalog and are thus omitted. The full list of LAT-detected pulsars is regularly updated,\footnote{See \url{https://confluence.slac.stanford.edu/display/GLAMCOG/Public+List+of+LAT-Detected+Gamma-Ray+Pulsars}.} and the current population will be discussed  in the forthcoming third LAT pulsar catalog.  We also point out that associations with pulsar candidates are based on spatial coincidence only and that $\gamma$-ray pulsations from these sources are not necessarily expected to be detected soon.

\subsection{Census of new DR2/DR3 sources relative to 4FGL-DR1}

For completeness, we recapitulate below the census of associations of new DR2 sources already given in the DR2 document \citep{LAT20_4FGLDR2}, with some updates on blazar classes:
\begin{enumerate}
\item two pulsar candidates, PSR J1439$-$5501 and PSR J1904$-$11;
\item one globular cluster (GLC), NGC 362;
\item one high-mass binary (HMB), PSR B1259$-$63; although detected by the LAT during its 2010 periastron passage \citep{LAT11_PSRB1259}, this source is included in an FGL catalog for the first time thanks to the gain in significance from its 2017 passage \citep{Assoc18_Joh};
\item three SNRs, 3C 397, SNR G001.4$-$00.1, and SNR G003.7$-$00.2; they are considered as reliable associations (as opposed to SPPs, see below) because their radio sizes are quite small ($4\arcmin$--$14\arcmin$) and their positions well match those of the 4FGL sources;
\item two star-forming regions, NGC 346, which is the brightest star-forming region in  the SMC, and Sh 2$-$152;
\item two galaxies, IC 678 and NGC 5380;
\item one starburst galaxy, Arp 299;
\item two radio galaxies, NGC 3078 and NGC 4261; 
\item 281 blazars, including 176 blazars of unknown type (BCUs), 66 BL Lac objects (BLL), and 39 FSRQs;\footnote{8 BCUs have been classified into 7 BLL and one FSRQ since the DR2 release}
\item 16 SPPs (SPP stands for ``SNR, Pulsar or PWN'' and refers to sources of unknown nature but overlapping with known SNRs or PWNe and thus candidate members of these classes);
\item 30 sources of unknown nature (UNK, $|b|<10\arcdeg$ sources solely associated with the likelihood-ratio method from large radio and X-ray surveys).
\end{enumerate}
The pulsar PSR J1757$-$2421, formerly associated with 4FGL J1756.6$-$2352 in DR1, is now associated with 4FGL J1757.9$-$2419. The source 4FGL J1756.6$-$2352 is now a SPP.

Below is a census of the associations for the new DR3 sources:  
\begin{enumerate}
\item three pulsars with detected pulsations, including two young pulsars, PSR J0922+0638 and PSR J1731$-$4744 \citep{Smi19}, and one MSP, PSR J1455$-$3330;
\item one MSP candidate, PSR J1957+2516; 
\item five GLCs, NGC 1851, NGC 5286, NGC 6205, NGC 6712, and M 54; 
\item two PWNe, PWN G327.1$-$1.1 and PWN G54.1+0.3;
\item one SNR, the Kepler SNR; the DR3 source 4FGL J1730.4$-$2131 was associated by the Likelihood Ratio method with NVSS J173036$-$212910; this NVSS source is actually part of the Kepler SNR (extended for the Very Large Array at 3$\arcmin$ diameter), so we report the SNR as the counterpart \citep[see also][]{Kepler21};
\item three novae, V1369 Cen \citep{Cheung_16}, V906 Car \citep{Jean_18, V906Car_Aydi20}, and YZ Ret \citep[MGAB-V207;][]{Li20}; the position of the DR3 source (4FGL J0358.5$-$5432) is closer to that of a blazar candidate, PMN J0358$-$5434, than to that of YZ Ret, but the light curve clearly points to the nova association;
\item one high-mass X-ray binary, 1RXS J172006.1$-$311702 \citep{Esp14}; the nature of this source is debated as it was classified as a cataclysmic variable in \cite{For18};
\item one radio galaxy, LEDA 55267, classified as a Fanaroff-Riley type 0 \citep{Bal19} with a LAT detection reported in \citet{Pal21}, and  one AGN, NGC 6454;
\item 306 blazar candidates including 219 BCUs, 51 BLLs, and 36 FSRQs; 
\item 22 SPPs;
\item 20 sources of unknown nature.
\end{enumerate}

\subsection{Association/classification changes in 4FGL-DR1 sources}

\begin{enumerate}
\item A specific class (GC) has been created for the Galactic center (4FGL J1745.6$-$2859), replacing its former SPP class.
\item The increase of localization systematic uncertainties for sources lying at $ |b|<5\arcdeg$ (\S~\ref{catalog_detection}) has led to the association of five extra BCUs, five SPPs, and the suppression of the association of 4FGL J1804.9$-$3001  with the GLC NGC 6528. Four sources classified as unknown in DR1  are now classified as BCUs because they gained Bayesian-based associations. Two other formerly unknown sources are now associated with a pulsar and a binary star (Kleinmann's star).
\item Two associations with pulsar candidates for 4FGL J1618.7$-$4633  and 4FGL J1720.8$-$1937 have been discarded on the realization that their estimated $\gamma$-ray luminosities would exceed their spin-down powers by factors greater than 30.
\item Pulsations have been detected for 22 more DR1 sources, including 10 PSR and 12 MSP \citep[e.g.,][]{Cla21,Tab21}. 
\item  A total of 29 associations with MSP from the West Virginia University list\footnote{\url{http://astro.phys.wvu.edu/GalacticMSPs/GalacticMSPs.txt}} have been added (three replacing associations with BCUs and one with a SPP), along with one association with a young pulsar candidate.
\item The pulsar PSR J1909$-$3744 was mistakenly associated with 4FGL J1912.2$-$3636. The position of this LAT-detected pulsar was used as a seed, but the resulting 4FGL source ended up too far away from the seed to make the association plausible. The 4FGL source is now unassociated.
\item The association of the MSP binary recently discovered by \citet{Assoc20_Wang} with  4FGL J0935.3+0901 has been implemented. Its class is binary (BIN).
\item Following \citet{Assoc15_Bog}, 4FGL J1544.5$-$1126 is a candidate transitional MSP binary (1RXS J154439.4$-$112820). Its class, BCU in DR1, has been changed to ``low-mass X-ray binary'' (LMB). Another candidate transitional MSP binary, CXOU J110926.4$-$650224, was found \citep{Zel19,Zel21} in a systematic search for X-ray counterparts to unassociated sources in the preliminary LAT 8 yr point-source list (FL8Y). While the position of the X-ray source lies well within the 95\% error ellipse of the FL8Y source, it is outside of the ellipse of the corresponding DR3 source 4FGL J1110.3$-$6501. In consequence,\footnote{Because of the targeted search in the X-rays, applying the standard association procedure would be inappropriate in this case.} we have not retained this association in our tables.  
\item Another transitional MSP binary, 1SXPS J042749.2$-$670434, displaying simultaneous optical, X-ray, and $\gamma$-ray  eclipses has been identified by \citet{Assoc20_Ken} in 4FGL J0427.8$-$6704 (previously unassociated).
\item  Gamma$^2$ Velorum \citep[4FGL J0809.5$-$4714;][]{Marti_20} has been classified as a binary system.
\item  We have implemented the results of \citet{Cor_19}, who  discovered that 4FGL J1405.1$-$6119 is a HMB. 
\item The source 4FGL J1832.9$-$0913, associated with the TeV source HESS J1832$-$093, has been classified as a HMB \citep[see][]{Mar_20}.
\item \citet{Swi_21} found a candidate redback MSP binary in a follow-up observation of 4FGL J0940.3$-$7610. 
Other redbacks have been discovered in a similar fashion in 4FGL J0407.7$-$5702 \citep{Mil20} and 4FGL J0540.0$-$7552 \citep{Stra21}. These associations have been classified as LMBs.
\item Similarly, \citet{Li_21} found a black widow binary at the position of 4FGL J0336.0+7502, which has then also been classified as a LMB. 
\item Three periodic optical and X-ray sources were recently discovered positionally consistent with DR1 sources. These redback MSP candidates have been classified as BIN. They are 4FGL J0212.1+5321 \citep{J0212_Li16}, 4FGL J0523.3$-$2527 \citep{J0523_Strader14}, and 4FGL J0838.7$-$2827 \citep{J0838_Halpern17}.
\item The tentative association of 4FGL J0647.7$-$4418 with the HMB  RX J0648.0$-$4418 called out in DR1 has been replaced by association with the BCU SUMSS J064744$-$441946 following the multiwavelength investigation of  \citet{Assoc20_Marti}.
\item Following \citet{Jar_20}, we have reclassified TXS 2116$-$077 (4FGL J2118.8$-$0723) as a Seyfert galaxy instead of a NLSY1. 
\item The latest version of the Radio Fundamental Catalog\footnote{rfc\_2021a available at \url{http://astrogeo.org/rfc/}} has enabled the association of six  previously unassociated sources and two SPPs with  blazar candidates.
\item Recent follow-up observations of 4FGL blazars \citep{Fup_Pen17,Fup_Pen19,Pen20,Pen21a,Pen21b,Pen21c,Des19,Raj21} have enabled the classification of 240 former BCUs, 2 AGNs, and 2 UNKs into 214 BLLs and 30 FSRQs.
\item Two FSRQs (PKS 0736$-$770, TXS 1530$-$131)  have been reclassified as BCUs and one other (Two Micron All Sky Survey, 2MASS J02212698+2514338) as a BLL.
\end{enumerate}

Note: A typo has been found in \S~6.2 of the 4FGL paper concerning the name of the star cluster associated with the extended H\,{\sc ii} region encompassing  4FGL J1115.1$-$6118. The star cluster is NGC 3603 (and not NGC 4603).

Table \ref{tab:classes} lists the class tallies of the whole DR3 catalog.

\begin{deluxetable*}{lcrcr}
\setlength{\tabcolsep}{0.04in}
\tablewidth{0pt}
\tabletypesize{\scriptsize}
\tablecaption{LAT 4FGL-DR3 Source Classes 
\label{tab:classes}
}
\tablehead{
\colhead{Description} & 
\multicolumn{2}{c}{Identified} &
\multicolumn{2}{c}{Associated} \\
& 
\colhead{Designator} &
\colhead{Number} &
\colhead{Designator} &
\colhead{Number}
}
\startdata
Galactic center & GC & 1 &  \nodata & \nodata \\
Young pulsars, identified by pulsations & PSR & 135 & \nodata & \nodata \\
Young pulsars, no pulsations seen in LAT yet & \nodata & \nodata & psr & 2 \\
Millisecond pulsars, identified by pulsations & MSP & 120 & \nodata & \nodata \\
Millisecond pulsars, no pulsations seen in LAT yet & \nodata & \nodata & msp & 35 \\
Pulsar wind nebula & PWN & 11 & pwn & 8 \\
Supernova remnant & SNR & 24 & snr &  19 \\
Supernova remnant / Pulsar wind nebula & SPP &  0  & spp  & 114 \\
Globular cluster & GLC &  0  & glc & 35 \\
Star-forming region & SFR & 3 & sfr &  2  \\
High-mass binary & HMB & 8 & hmb & 3 \\
Low-mass binary & LMB & 2 & lmb & 6 \\
Binary & BIN & 1 & bin & 6 \\
Nova & NOV & 4 & nov & 0  \\
BL Lac type of blazar & BLL & 22 & bll & 1435 \\ 
FSRQ type of blazar &   FSRQ & 44 & fsrq & 750 \\
Radio galaxy & RDG & 6 & rdg & 39 \\
Nonblazar active galaxy & AGN & 1 & agn & 8\\ 
Steep spectrum radio quasar & SSRQ & 0 & ssrq & 2 \\
Compact steep spectrum radio source & CSS &  0  & css & 5 \\
Blazar candidate of uncertain type & BCU & 1 & bcu & 1491 \\
Narrow-line Seyfert 1 & NLSY1 & 4 & nlsy1 & 4 \\
Seyfert galaxy & SEY &  0  & sey & 2 \\
Starburst galaxy & SBG &  0  & sbg & 8 \\
Normal galaxy (or part) & GAL & 2 & gal & 4 \\
Unknown & UNK & 0 & unk & 134 \\
Total & \nodata &  389 & \nodata &  4112 \\
\hline
Unassociated & \nodata & \nodata & \nodata &  2157 \\ 
\enddata
\tablecomments{The designation `spp' indicates potential association with SNR or PWN. `Unknown' are $|b| < 10\arcdeg$ sources solely associated with the likelihood-ratio method from large radio and X-ray surveys. Designations shown in capital letters are firm identifications; lower-case letters indicate associations. }
\end{deluxetable*}

\subsection{Spectral properties}

\begin{figure}[!ht]
\centering
\includegraphics[width=0.5\textwidth]{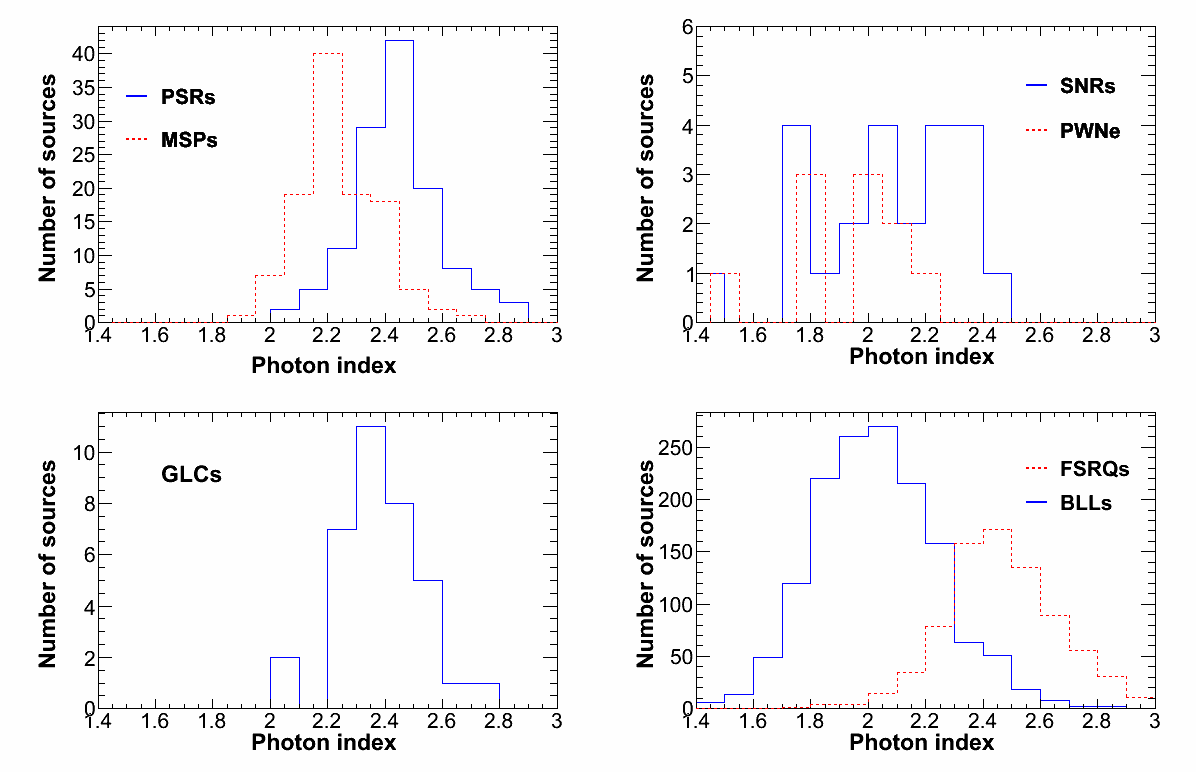}
\caption{Photon-index distributions for different source classes. }
\label{fig:index_class}
\end{figure}

In this section, the spectral properties for the main source classes in the whole 4FGL-DR3 sample are presented. We use these properties to shed light on the nature of the unassociated sources in the next section.   Although some classes, most notably pulsars, show significant levels of spectral curvature, the PL photon index, $\Gamma$, is a convenient parameter enabling comparison across classes. Figure \ref{fig:index_class} compares the photon-index distributions for different established classes. Pulsars show remarkably consistent soft spectra, along with GLCs. MSP have somewhat harder spectra on average compared to PSR and GLCs.  The observed difference between the MSP and GLC distributions seems to challenge the conventional wisdom that MSP dominate the GLC's $\gamma$-ray emission \citep[e.g.,][]{Har05,Ven09}, and is possibly due to different ambient stellar radiation fields comptonized by the MSP electrons \citep{MSP_Bend07}. The spectra of SNRs and PWNe are notably harder than those of pulsars, in agreement with the observation that several SNRs and almost all PWNe are detected in the TeV domain. The distribution for blazars spans a broad range of photon index, the distributions for FSRQs and BLLs exhibiting moderate overlap \citep[see][]{LAT20_4LAC}. Blazars constitute the only abundant class showing a hard-spectrum component. 
     
\begin{figure}[!ht]
\centering
\includegraphics[width=0.5\textwidth]{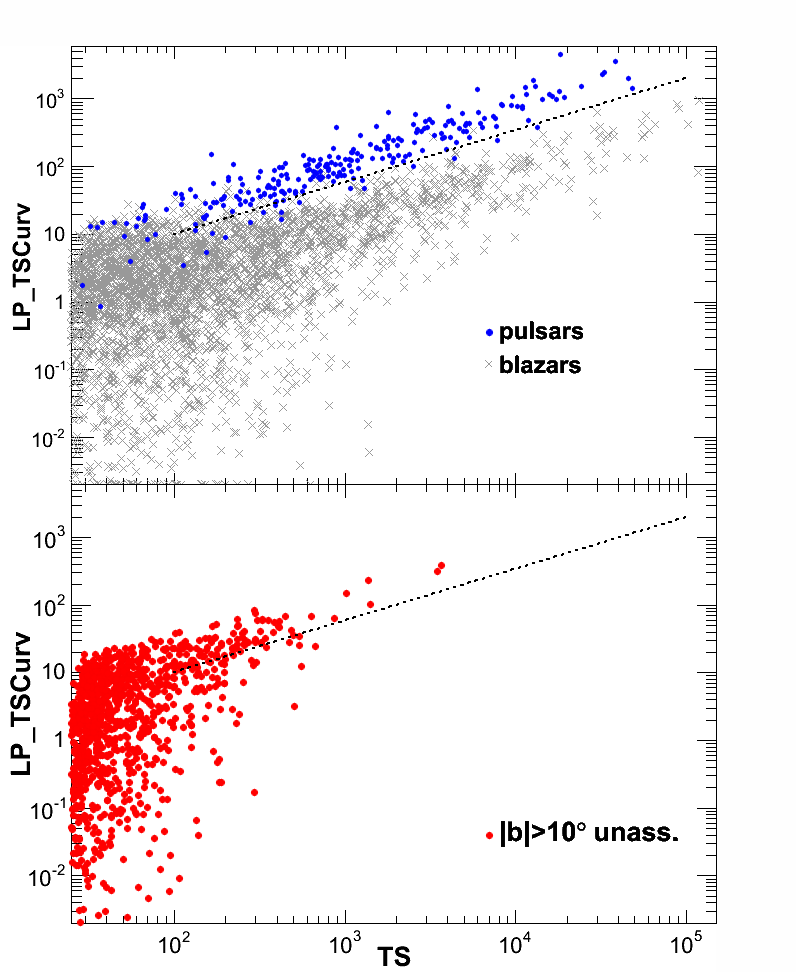}
\caption{Top: parameter LP\_TS$_{\rm curv}$ (reflecting the spectral-curvature significance) as a function of Test Statistic for blazars and pulsars (no selection on the Galactic latitude was applied). The dotted line, selected by eye, approximately delineates the separation between the two corresponding branches.  Bottom: same for  the high-latitude unassociated sources. The same dotted line as in the top panel is displayed for orientation.}
\label{fig:TS_curv}
\end{figure}

Turning to the spectral curvature, while some blazars show significant curvature, as do  most pulsars, the two  classes occupy different branches in the LP\_TS$_{\rm curv}$ versus TS plane (Figure \ref{fig:TS_curv} top). These branches merge at the low-TS end as curvature cannot be assessed with high significance in that case. Despite the fact that PSR and MSP show slightly different spectra as can be seen from their photon-index distributions, both fall on the same branch in Figure \ref{fig:TS_curv} top.

\begin{figure}[!ht]
\centering
\includegraphics[width=0.5\textwidth]{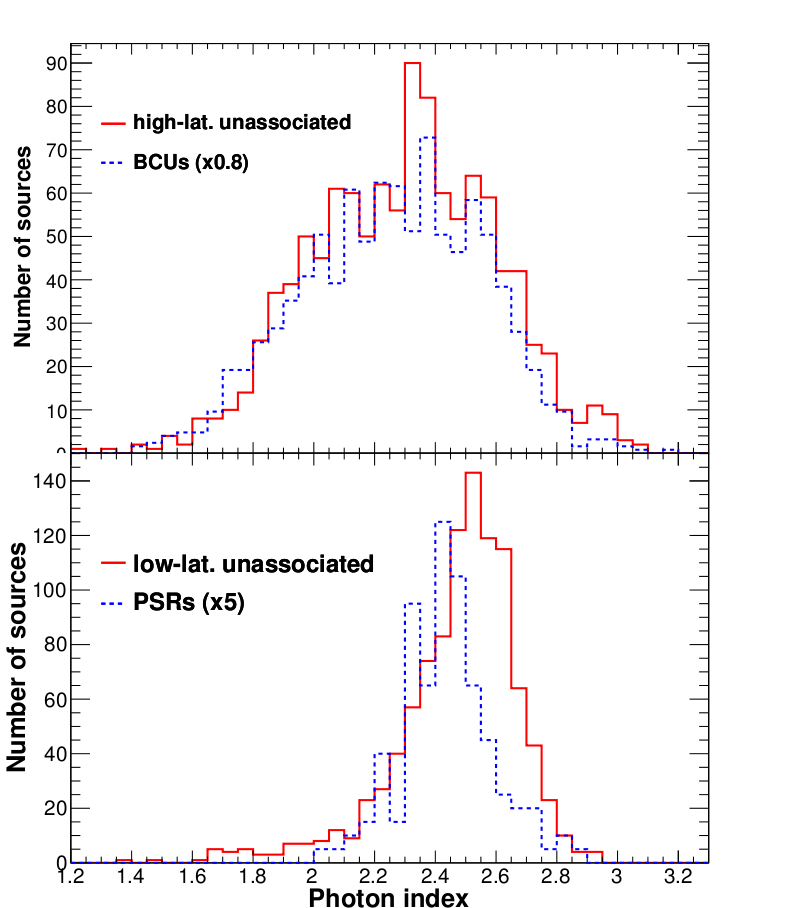}
\caption{Top: photon-index distribution of high-latitude unassociated sources compared to that of blazars of unknown type. Bottom: photon-index distribution of low-latitude unassociated sources compared to that of young pulsars. }
\label{fig:unass_index}
\end{figure}

\begin{figure}[!ht]
\centering
\includegraphics[width=0.5\textwidth]{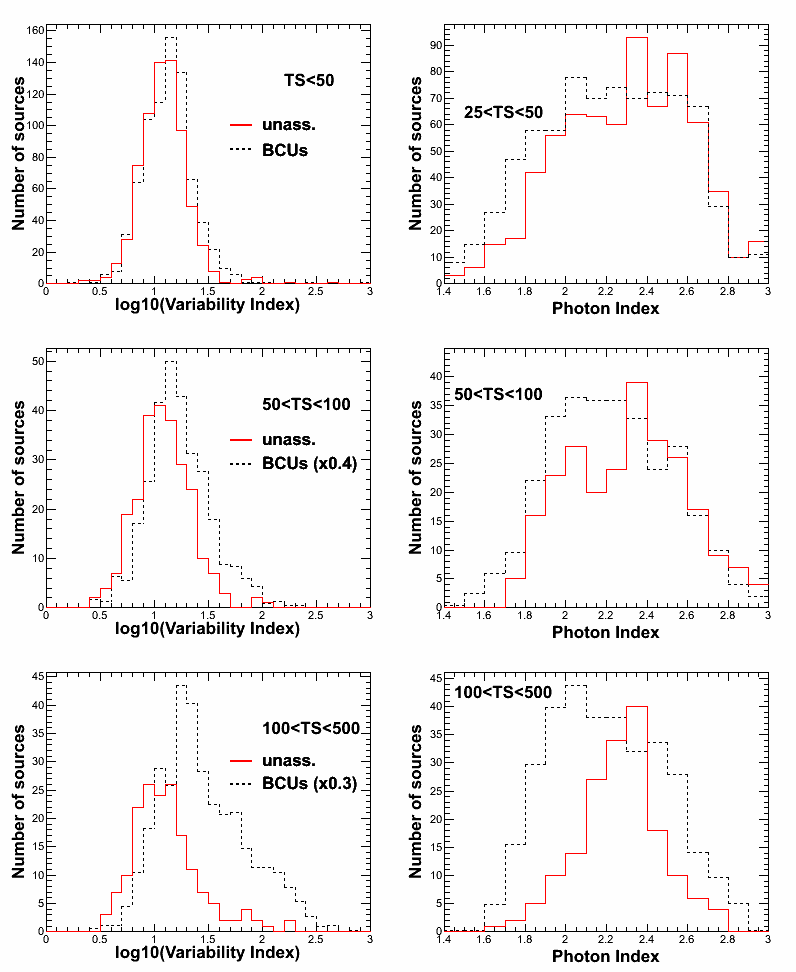}
\caption{Variability-index (left) and photon-index (right) distributions of high-latitude unassociated sources and blazars for different bins in Test\_Statistic.}
\label{fig:var_index}
\end{figure}

\section{Unassociated sources}
\label{dr3_unassocs}

The variation of the photon-index distribution of unassociated sources with Galactic latitude indicates that different populations coexist (see Figure 21 of the 4FGL paper), with the spectra becoming softer on average for lower Galactic latitudes. This observation justifies treating separately the cases of high- and low-latitude unassociated sources. We set a somewhat arbitrary limit at $|b|=10\arcdeg$, with 1122 and 1018 unassociated point sources lying at  higher and lower latitudes respectively.    

\subsection{High-latitude unassociated sources}

Drawing on the class-wise discriminating power of the photon-index distributions illustrated in  Figure \ref{fig:index_class}, the distribution for high-latitude unassociated sources is compared to that of BCUs, which are the most relevant blazar set given the low TS, in  Figure \ref{fig:unass_index} top.  The good agreement between the two distributions supports the conclusion that most high-latitude unassociated sources are indeed blazars\footnote{No other source class shows a similar distribution; see Figure \ref{fig:index_class}.}. This conclusion is slightly qualified when considering the loci of the unassociated sources in the TS versus LP\_TS$_{\rm curv}$ plane (Figure \ref{fig:TS_curv} bottom).  A significant fraction ($\simeq$70\%) of the higher-TS (TS $>$ 100) unassociated sources fall onto the pulsar branch, strongly suggesting that they are MSP. This idea is also supported by considering the evolution of the photon-index and  variability-index  distributions with TS (Figure \ref{fig:var_index}). While for the lower-TS bins, the distributions quite closely match those of BCUs, clear differences are observed for the higher-TS bin (100 $<$ TS $<$ 500), where the distributions are compatible with those expected for MSP. In conclusion, their photon indices,  significance of spectral curvature, and lack of variability make many of the high-latitude, 100 $<$ TS $<$ 500 unassociated sources good MSP candidates.

\subsection{Low-latitude unassociated sources}
\label{lowlatunassocs}

 Since some AGN catalogs suffer from significant incompleteness close to the plane due to extinction, some background blazars detected through the plane could elude association and contribute to the unassociated sample. Based on the tallies observed at high latitude and the difference in flux limits due to the brighter diffuse emission background close to the plane, the number of  $|b|<10\arcdeg$ blazars is estimated to be 340$\pm$20, while 399 blazars are actually reported as associations in DR3. Comparing the low-latitude photon-index distribution with that obtained for higher latitudes (normalized so that the total yield of  $\Gamma<$ 2.4 sources match), an excess of soft sources ($\Gamma\simeq$ 2.5) is found in the low-latitude sample relative to the other. From the difference of the spectral-index distributions, this excess, which was less pronounced in 3FGL and  is very likely due to a contamination from a nonblazar population, amounts to 75$\pm$4 sources. In conclusion, an excess of soft-spectra sources stands out in the low-latitude blazar sample.  It can be attributed to a contamination from another population, likely of Galactic origin. After subtracting this excess from the number of reported blazars, the remaining number of detected blazars agrees reasonably well with that extrapolated from the high-latitude population (324 vs. 340$\pm$20). Few blazars with missed associations are thus expected to contribute to the low-latitude unassociated sample.            
\begin{figure}[!ht]
\centering
\includegraphics[width=0.5\textwidth]{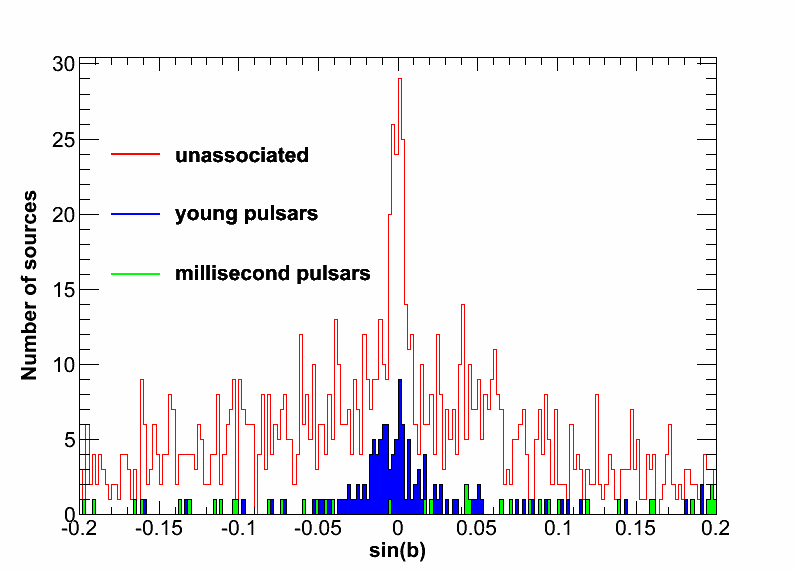}
\caption{Galactic-latitude distribution of unassociated sources around the Galactic plane compared to that of pulsars.}
\label{fig:lat_dist}
\end{figure}

\begin{figure*}[!ht]
\centering
\includegraphics[width=1.0\textwidth]{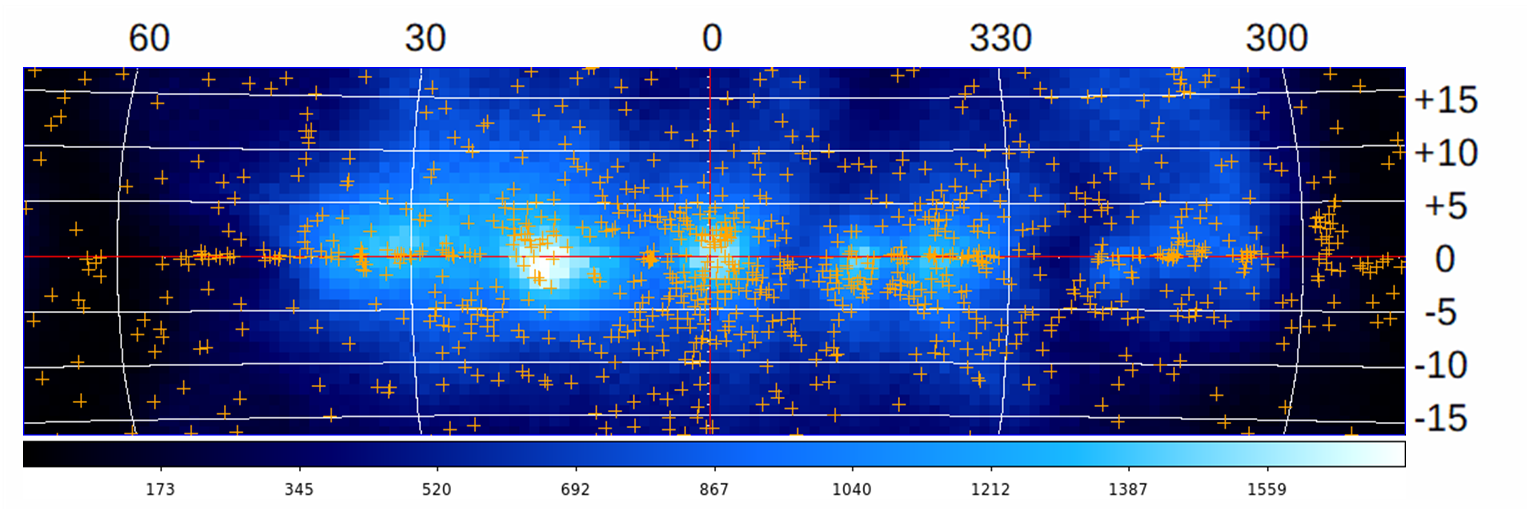}
\caption{Positions of unassociated sources (crosses) around the Galactic center. The  background is a count map of 83--228 MeV $patch$ photons simulated  over 12 yr with a pixel size of 1$\arcdeg$.}
\label{fig:sgu_map}
\end{figure*}

Unassociated sources around the Galactic plane (referred to as GUs) account for a growing fraction of the detected sources, from 12\% in DR1 to 24\% in the new sample. The features presented here were already present in DR1 but have become more salient with the larger population.

The GU latitude distribution, shown in Figure \ref{fig:lat_dist}, shows a narrow Gaussian-like component ($\simeq$ 210 sources) centered on the plane,  and two broad shoulders ($\simeq$ 760 sources)  extending out to about $|b|=10\arcdeg$ (these components will be referred to as ``spike" and ``shoulders" respectively). The spike is narrower ($\sigma_{\sin(b)}$=0.005) than the young-pulsar distribution ($\sigma_{\sin(b)}$=0.012 for sources with $\sin(b)<0.03$). The loci of the unassociated sources close to the plane exhibit a notable clustering. Some high-density regions (``hot spots") are readily seen in  Figure \ref{fig:sgu_map},  while other areas are mostly devoid of unassociated sources. The clustering aspect also manifests itself in the large fraction (30\%) of GUs with Flag 5 (i.e., close to a brighter neighbor). High-density regions have been determined using a HEALPix tesselation with $N_{\rm side}=8$ produced in Galactic coordinates. The average density of GUs per pixel in the three pixel rows spanning  the Galactic Plane is about 7. We define the high-density pixels as those encompassing more than 14 GUs, i.e. twice the average value. 
   
\begin{figure}[!ht]
\centering
\includegraphics[width=0.5\textwidth]{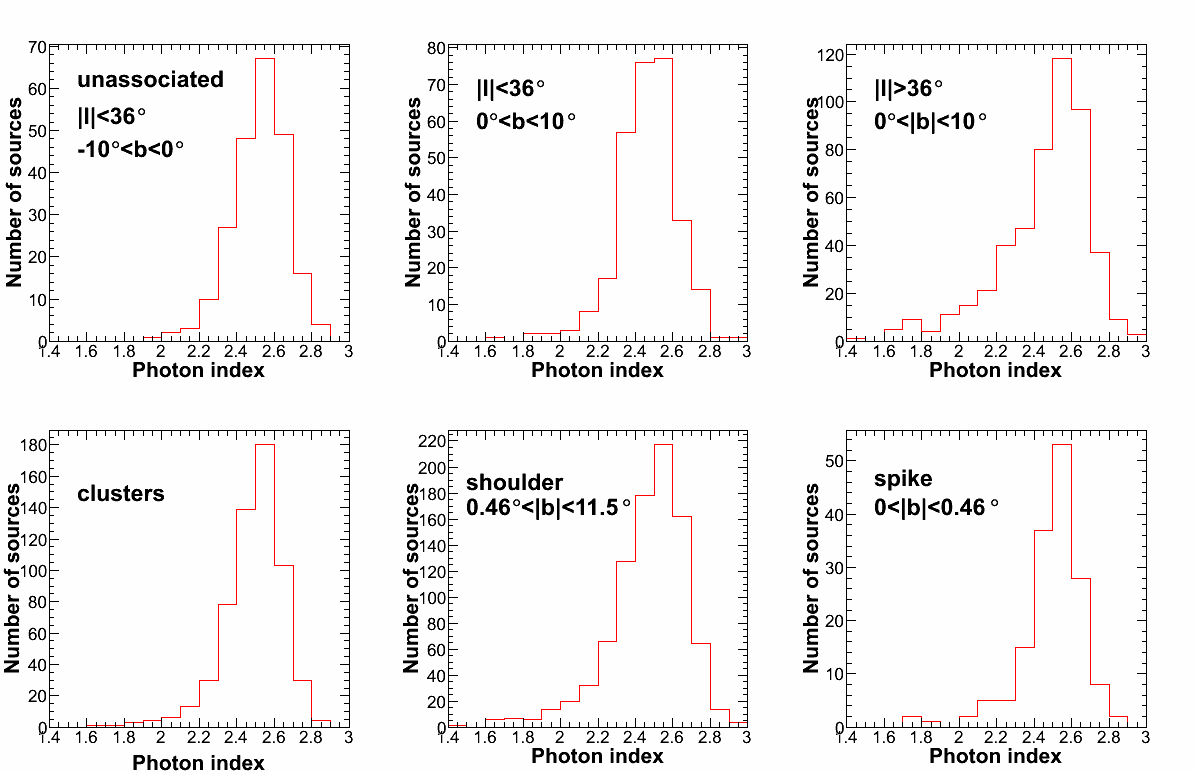}
\caption{Photon-index distributions of low-latitude unassociated sources in different sky regions. }
\label{fig:index_regions}
\end{figure} 

The photon-index distribution is compared to that of PSR in the lower panel of  Figure \ref{fig:unass_index}. The distribution is remarkably narrow and contrasts sharply with that observed at higher latitude  (Figure \ref{fig:unass_index} top). It peaks around $\Gamma=2.5$, with a tail toward lower $\Gamma$ values. This tail might be due to blazars, SNRs, or PWNe (Figure \ref{fig:index_class}). The high-$\Gamma$ end of the distribution is peculiar. No established class of Galactic $\gamma$-ray emitters exhibits such a soft component.  We will refer to sources in this soft component as SGUs.

A dedicated flag (Flag 14) has been devised to indicate SGUs  that are in regions of relatively high source density as seen in Figure \ref{fig:sgu_map}. 
 Flag 14 is set for SGUs lying inside a high-density pixel, having $\Gamma>2.4$, TS $<$ 500, and with a closest neighbor (classified  either as GU, SPP, or UNK) less than 2$\arcdeg$ away. Since most SPP or UNK sources probably belong to the same population as GUs as ascertained by their similar photon-index distributions (peaking around $\Gamma$=2.5),  they are also flagged if the above conditions are filled. Flag 14 is set for 446 SGUs, 54 SPPs, and 49 UNKs.\footnote{We checked that the set of flagged sources depends moderately (typical $>$ 85\% overlap) on the centering of the HEALPix pixels.} SGUs were already significantly present in DR1, but their number in high-density regions has now increased by 50\% (see census of Flag 14 in Table~\ref{tab:flags}).

\begin{figure}[!ht]
\centering
\includegraphics[width=0.5\textwidth]{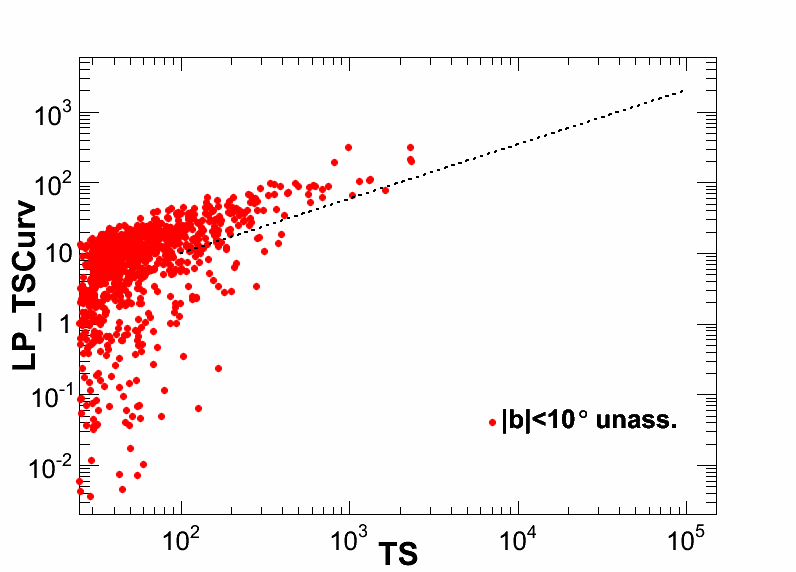}
\caption{Same as Figure \ref{fig:TS_curv} bottom, but for low-latitude unassociated sources.}
\label{fig:TS_curv_l}
\end{figure}

Returning to the spectral properties, the photon-index distribution is essentially  constant across the plane, as illustrated in Figure \ref{fig:index_regions} for different (not mutually exclusive) sky regions. In this figure, ``clusters" refers to the high-density pixels mentioned above.  The unassociated sources  preferably lie within the pulsar branch in the LP\_TS$_{\rm curv}$ vs. TS plane (Figure \ref{fig:TS_curv_l}), supporting their spectral resemblance with pulsars.  Their distribution of the LP curvature parameter $\beta$ is also consistent with that of pulsars. Despite these similarities, other factors distinguish  GUs and pulsars.
First, the photon-index distributions do not closely match. The significant clustering observed also disfavors the possibility that GUs are mostly pulsars. Moreover, the low-$b$ unassociated sources  outnumber the detected PSR by about a factor of 9, making the scenario that so many pulsars have gone undetected so far implausible. The final blow to this possibility comes from their broad latitude distribution, which does not match that of PSR (nor any of the distributions of known $\gamma$-ray emitters for that matter).
Older pulsars ($>10^6$ yr), which could have  drifted farther off the plane over time, exhibit a wider Galactic-latitude distribution, more compatible with that observed. However,  these pulsars should be undetectable in $\gamma$-rays as they would have crossed the `pair-creation death line' \citep[see][]{LAT13_2PC}.

Most GUs exhibit a low source-to-background ratio (30\% of them having  the corresponding Flag 4 set). Whether they originate from mismodeled Galactic diffuse emission becomes a valid question. This emission is not fully understood, and its modeling  suffers many uncertainties. To produce the Pass 8 model\footnote{\url{https://fermi.gsfc.nasa.gov/ssc/data/analysis/software/aux/4fgl/Galactic_Diffuse_Emission_Model_for_the_4FGL_Catalog_Analysis.pdf}},  
the contributions of different components, H {\textsc i}, CO, dark neutral medium, and inverse Compton,  whose properties are established at larger wavelengths,  were fitted to the data in 9 rings of increasing Galactocentric distances. An ad hoc component called the patch encapsulates the contributions of nontemplate components like the Galactic center excess, Loop I, or the Fermi bubbles. Other, yet-to-be-identified large scale structures may be contributing \citep[see][for a search performed at high latitude]{Gulli2021}, and a comprehensive analysis of the patch  still needs to be carried out. Understanding the patch low-energy behavior would be of particular relevance to this work. The patch covers a limited sky area and was smoothed out not to absorb features with spatial scales less than 4$\arcdeg$. Looking for spatial correlations between the SGUs and features present in the different components, the best match was obtained with the patch component at low energy (see Figure \ref{fig:sgu_map}). The SGU latitude distribution exhibits a north/south symmetry and a relative extension (Figure \ref{fig:lat_dist}) that find a good correspondence in the patch at $|b|<10\arcdeg$.  Comprehending the photon excess underlying the patch will certainly be key in unveiling the nature of the SGUs.  

Below we list some possibilities regarding the SGUs and lay out avenues for further exploring them.

\begin{itemize}
\item As mentioned above, SGUs may result from mismodeled Galactic diffuse emission. 
Allowing more flexibility to the model fitting by leaving free individual diffuse components, like H {\textsc i} and CO, would constrain the robustness of the SGU detections, if the templates used for the model are accurate on scales that are finer than the RoI size, while inaccurate on larger scales.

\item A  possibility to be considered is that  SGUs trace the large-scale regions associated with  fresh injection of cosmic-rays, leading to locally harder $\gamma$-ray emission than assumed in the model components. A worthwhile test would be to investigate  (e.g., by means of simulations) whether the spurious point-like sources spawned from a possible, not-accounted-for diffuse excess would share common spectral features with the observed SGUs. 

\item Some SGUs could be spatially extended if they are related to structures in the diffuse emission. This  would introduce some  spectral distortion when analyzing them under the assumption that they are point-like, making them softer than they really are.   

\item Some correlation with the general structures of interstellar clouds is  found by visual inspection, suggesting that  the emission ascribed to SGUs could be due to molecular gas unaccounted for on small scales, e.g., because of saturated CO lines. This situation is reminiscent of that encountered in the EGRET era, which led to the discovery of ``dark gas'' \citep{EGRET_DarkGas} and the suppression of about 30\% of unassociated EGRET sources \citep{EGRcatalog} once a modified diffuse emission model was considered. Assessing the effect of saturated CO lines using $^{13}$CO data would shed light on this possibility. 

\item Alternative ways could be pursued to build the patch (diffuse emission that is not accounted for by standard components of the interstellar emission model) and distinguish it from point sources \citep[e.g.,][]{D3PO15}.

\item The alternative to mismodeled diffuse emission,  namely the existence of an abundant, entirely new class of sources remains highly speculative. Follow-up multiwavelength studies \citep[as in][]{Bru21}, e.g., of regions of high SGU densities or surrounding the brightest SGUs, could help elucidate this issue.
\end{itemize}

\section{Conclusions}
\label{conclusions}

We built two incremental versions (DR2 after 10 yr and DR3 after 12 yr) of the fourth \Fermilat source catalog, the deepest-yet in the GeV energy range.
Incremental versions preserve the existence and positions of sources in the previous versions (with a few exceptions, see \S~\ref{catalog_detection}), and add new sources.
We will continue providing incremental versions every two years, until a major ingredient (the event reconstruction, the calibrations, or the underlying diffuse emission model) improves significantly.

We applied a new method to check the quality of the global model of the sky (see \S~\ref{catalog_catxcheck}). That method indicates that the interstellar emission model is too bright in a number of local molecular clouds.
Conversely, it may be too faint close to the Galactic plane where we find many clustered soft unassociated sources (see \S~\ref{lowlatunassocs}).
Understanding the nature of these sources and what fraction are actually diffuse emission will be an important challenge, which could potentially lead to major advances like the discovery of the dark gas in the EGRET data \citep{EGRET_DarkGas}.

The 4FGL-DR3 catalog includes 6658 sources, 1695 of which (25.5\%) are found to be significantly variable on one-year timescales, and 3388 of which (50.9\%) are fit with curved spectral shapes (see \S~\ref{catalog_spectra}). We mark 317 (4.8\%) of the sources as potentially related to imperfections in the model for Galactic diffuse emission; the character \texttt{c} (for ``on top of interstellar gas clump'') is appended to their names (except those already marked as \texttt{e} for extended).  An additional 1836 (27.6\%) are flagged in the catalog for less serious concerns, e.g., for depending sensitively on the details of the analysis or for being close to a brighter source (see \S~\ref{catalog_analysis_flags}).

Of the 6658 sources in the catalog, 389 (5.8\%) are considered identified, based on pulsations, correlated variability, or correlated angular sizes with observations at other wavelengths. We find likely lower-energy counterparts for 4112 other sources (61.8\%).  The remaining 2157 sources (32.4\%) are unassociated.

The identified and associated sources in the 4FGL-DR3 catalog include many Galactic and extragalactic source classes (see \S~\ref{dr3_assocs}).  The largest Galactic source class continues to be pulsars, with 135 young, 120 millisecond $\gamma$-ray pulsars, and 37 associations to non-LAT pulsars (mostly MSP). Other Galactic source classes have continued to grow; 35 GLCs, 43 SNRs and 19 PWNe are now associated with LAT sources.  Blazars remain the largest class of extragalactic sources, with 2251 identified or associated with BL Lac objects or FSRQ active galaxies and 1492 with unclassified blazars.

\begin{acknowledgments}
The \textit{Fermi} LAT Collaboration acknowledges generous ongoing support
from a number of agencies and institutes that have supported both the
development and the operation of the LAT as well as scientific data analysis.
These include the National Aeronautics and Space Administration and the
Department of Energy in the United States, the Commissariat \`a l'Energie Atomique
and the Centre National de la Recherche Scientifique / Institut National de Physique
Nucl\'eaire et de Physique des Particules in France, the Agenzia Spaziale Italiana
and the Istituto Nazionale di Fisica Nucleare in Italy, the Ministry of Education,
Culture, Sports, Science and Technology (MEXT), High Energy Accelerator Research
Organization (KEK) and Japan Aerospace Exploration Agency (JAXA) in Japan, and
the K.~A.~Wallenberg Foundation, the Swedish Research Council and the
Swedish National Space Board in Sweden.
 
Additional support for science analysis during the operations phase is gratefully
acknowledged from the Istituto Nazionale di Astrofisica in Italy and the Centre
National d'\'Etudes Spatiales in France. This work performed in part under DOE
Contract DE-AC02-76SF00515.

This work made extensive use of the ATNF pulsar  catalog\footnote{\url{http://www.atnf.csiro.au/research/pulsar/psrcat}}  \citep{ATNFcatalog}.  This research has made use of the \citet{NED1} which is operated by the Jet Propulsion Laboratory, California Institute of Technology, under contract with the National Aeronautics and Space Administration, and of archival data, software, and online services provided by the ASI Science Data Center (ASDC) operated by the Italian Space Agency.
We used the Manitoba SNR catalog \citep{Ferrand2012_SNRCat} to check recently published extended sources. We acknowledge the Einstein@Home project
for providing new pulsar associations through the dedicated efforts of the Einstein@Home volunteers. The Einstein@Home project is supported by the  NSF award 1816904.
\end{acknowledgments}

\software{Gardian \citep{Diffuse2}, GALPROP\footnote{\url{http://galprop.stanford.edu}} \citep{GALPROP17}, HEALPix\footnote{\url{http://healpix.jpl.nasa.gov/}} \citep{Gorski2005}, Aladin\footnote{http://aladin.u-strasbg.fr/}, TOPCAT\footnote{\url{http://www.star.bristol.ac.uk/\~mbt/topcat/}} \citep{Tay05}}

\facility{\Fermilat}

\bibliography{Bibtex_4FGL_v1}

\begin{thebibliography}{}
\expandafter\ifx\csname natexlab\endcsname\relax\def\natexlab#1{#1}\fi
\providecommand{\url}[1]{\href{#1}{#1}}
\providecommand{\dodoi}[1]{doi:~\href{http://doi.org/#1}{\nolinkurl{#1}}}
\providecommand{\doeprint}[1]{\href{http://ascl.net/#1}{\nolinkurl{http://ascl.net/#1}}}
\providecommand{\doarXiv}[1]{\href{https://arxiv.org/abs/#1}{\nolinkurl{https://arxiv.org/abs/#1}}}

\bibitem[{{Abdo} {et~al.}(2010){Abdo}, {Ackermann}, {Ajello},
  {et~al.}}]{LAT10_SMC}
{Abdo}, A.~A., {Ackermann}, M., {Ajello}, M., {et~al.} 2010, \aap, 523, A46,
  \dodoi{10.1051/0004-6361/201014855}

\bibitem[{{Abdo} {et~al.}(2011){Abdo}, {Ackermann}, {Ajello},
  {et~al.}}]{LAT11_PSRB1259}
---. 2011, \apjl, 736, L11, \dodoi{10.1088/2041-8205/736/1/L11}

\bibitem[{{Abdo} {et~al.}(2013){Abdo}, {Ajello}, {Allafort},
  {et~al.}}]{LAT13_2PC}
{Abdo}, A.~A., {Ajello}, M., {Allafort}, A., {et~al.} 2013, \apjs, 208, 17,
  \dodoi{10.1088/0067-0049/208/2/17}

\bibitem[{{Abdollahi} {et~al.}(2020){Abdollahi}, {Acero}, {Ackermann},
  {Ajello}, {Atwood}, {et~al.}}]{LAT20_4FGL}
{Abdollahi}, S., {Acero}, F., {Ackermann}, M., {et~al.} 2020, \apjs, 247, 33,
  \dodoi{10.3847/1538-4365/ab6bcb}

\bibitem[{{Abdollahi} {et~al.}(2017){Abdollahi}, {Ackermann}, {Ajello},
  {et~al.}}]{LAT17_FAVA}
{Abdollahi}, S., {Ackermann}, M., {Ajello}, M., {et~al.} 2017, \apj, 846, 34,
  \dodoi{10.3847/1538-4357/aa8092}

\bibitem[{{Acero} {et~al.}(2022){Acero}, {Lemoine-Goumard}, \&
  {Ballet}}]{Kepler21}
{Acero}, F., {Lemoine-Goumard}, M., \& {Ballet}, J. 2022, {\aap}, in press.
\newblock \doarXiv{2201.05567}

\bibitem[{{Ackermann} {et~al.}(2012){Ackermann}, {Ajello}, {Atwood},
  {et~al.}}]{Diffuse2}
{Ackermann}, M., {Ajello}, M., {Atwood}, W.~B., {et~al.} 2012, \apj, 750, 3,
  \dodoi{10.1088/0004-637X/750/1/3}

\bibitem[{{Ajello} {et~al.}(2020){Ajello}, {Angioni}, {Axelsson}, {Ballet},
  {Barbiellini}, {et~al.}}]{LAT20_4LAC}
{Ajello}, M., {Angioni}, R., {Axelsson}, M., {et~al.} 2020, \apj, 892, 105,
  \dodoi{10.3847/1538-4357/ab791e}

\bibitem[{{Ajello} {et~al.}(2017){Ajello}, {Atwood}, {Baldini},
  {et~al.}}]{LAT17_3FHL}
{Ajello}, M., {Atwood}, W.~B., {Baldini}, L., {et~al.} 2017, \apjs, 232, 18,
  \dodoi{10.3847/1538-4365/aa8221}

\bibitem[{{Ambrogi} {et~al.}(2019){Ambrogi}, {Zanin}, {Casanova}, {De O{\~n}a
  Wilhelmi}, {Peron}, \& {Aharonian}}]{HB21_Ambrogi19}
{Ambrogi}, L., {Zanin}, R., {Casanova}, S., {et~al.} 2019, \aap, 623, A86,
  \dodoi{10.1051/0004-6361/201833985}

\bibitem[{{Araya}(2020)}]{G279_Araya20}
{Araya}, M. 2020, \mnras, 492, 5980, \dodoi{10.1093/mnras/staa244}

\bibitem[{{Araya} {et~al.}(2019){Araya}, {Mitchell}, \&
  {Parsons}}]{HESSJ1825_Araya19}
{Araya}, M., {Mitchell}, A.~M.~W., \& {Parsons}, R.~D. 2019, \mnras, 485, 1001,
  \dodoi{10.1093/mnras/stz462}

\bibitem[{{Atwood} {et~al.}(2009){Atwood}, {Abdo}, {Ackermann},
  {et~al.}}]{LAT09_instrument}
{Atwood}, W.~B., {Abdo}, A.~A., {Ackermann}, M., {et~al.} 2009, \apj, 697,
  1071, \dodoi{10.1088/0004-637X/697/2/1071}

\bibitem[{{Atwood} {et~al.}(2013){Atwood}, {Albert}, {Baldini},
  {et~al.}}]{LAT13_P8}
{Atwood}, W.~B., {Albert}, A., {Baldini}, L., {et~al.} 2013, Fermi Symposium
  proceedings - eConf C121028.
\newblock \doarXiv{1303.3514}

\bibitem[{{Aydi} {et~al.}(2020){Aydi}, {Sokolovsky}, {Chomiuk},
  {et~al.}}]{V906Car_Aydi20}
{Aydi}, E., {Sokolovsky}, K.~V., {Chomiuk}, L., {et~al.} 2020, NatAs, 4, 776,
  \dodoi{10.1038/s41550-020-1070-y}

\bibitem[{{Baldi} {et~al.}(2019){Baldi}, {Capetti}, \& {Giovannini}}]{Bal19}
{Baldi}, R.~D., {Capetti}, A., \& {Giovannini}, G. 2019, \mnras, 482, 2294,
  \dodoi{10.1093/mnras/sty2703}

\bibitem[{{Baldini} {et~al.}(2021){Baldini}, {Ballet}, {Bastieri}, {Becerra
  Gonzalez}, {Bellazzini}, {et~al.}}]{2021_1FLT}
{Baldini}, L., {Ballet}, J., {Bastieri}, D., {et~al.} 2021, \apjs, 256, 13,
  \dodoi{10.3847/1538-4365/ac072a}

\bibitem[{{Ballet} {et~al.}(2020){Ballet}, {Burnett}, {Digel}, \&
  {Lott}}]{LAT20_4FGLDR2}
{Ballet}, J., {Burnett}, T.~H., {Digel}, S.~W., \& {Lott}, B. 2020, arXiv
  e-prints, arXiv:2005.11208.
\newblock \doarXiv{2005.11208}

\bibitem[{Bednarek \& Sitarek(2007)}]{MSP_Bend07}
Bednarek, W., \& Sitarek, J. 2007, Monthly Notices of the Royal Astronomical
  Society, 377, 920, \dodoi{10.1111/j.1365-2966.2007.11664.x}

\bibitem[{{Bogdanov} \& {Halpern}(2015)}]{Assoc15_Bog}
{Bogdanov}, S., \& {Halpern}, J.~P. 2015, \apjl, 803, L27,
  \dodoi{10.1088/2041-8205/803/2/L27}

\bibitem[{{Bruel}(2021)}]{PSmap}
{Bruel}, P. 2021, \aap, 656, A81, \dodoi{10.1051/0004-6361/202141553}

\bibitem[{{Bruel} {et~al.}(2018){Bruel}, {Burnett}, {Digel}, {Johannesson},
  {Omodei}, \& {Wood}}]{LAT18_P305}
{Bruel}, P., {Burnett}, T.~H., {Digel}, S.~W., {et~al.} 2018, $8^{\rm th}$
  Fermi Symposium.
\newblock \doarXiv{1810.11394}

\bibitem[{{Bruzewski} {et~al.}(2021){Bruzewski}, {Schinzel}, {Taylor}, \&
  {Petrov}}]{Bru21}
{Bruzewski}, S., {Schinzel}, F.~K., {Taylor}, G.~B., \& {Petrov}, L. 2021,
  \apj, 914, 42, \dodoi{10.3847/1538-4357/abf73b}

\bibitem[{{Casandjian} \& {Grenier}(2008)}]{EGRcatalog}
{Casandjian}, J.-M., \& {Grenier}, I.~A. 2008, \aap, 489, 849,
  \dodoi{10.1051/0004-6361:200809685}

\bibitem[{{Cheung} {et~al.}(2016){Cheung}, {Jean}, {Shore}, {Stawarz},
  {Corbet}, {Kn{\"o}dlseder}, {Starrfield}, {Wood}, {Desiante}, {Longo},
  {Pivato}, \& {Wood}}]{Cheung_16}
{Cheung}, C.~C., {Jean}, P., {Shore}, S.~N., {et~al.} 2016, \apj, 826, 142,
  \dodoi{10.3847/0004-637X/826/2/142}

\bibitem[{{Clark} {et~al.}(2021){Clark}, {Nieder}, {Voisin}, {Allen},
  {Aulbert}, {Behnke}, {Breton}, {Choquet}, {Corongiu}, {Dhillon},
  {Eggenstein}, {Fehrmann}, {Guillemot}, {Harding}, {Kennedy}, {Machenschalk},
  {Marsh}, {Mata S{\'a}nchez}, {Mignani}, {Stringer}, {Wadiasingh}, \&
  {Wu}}]{Cla21}
{Clark}, C.~J., {Nieder}, L., {Voisin}, G., {et~al.} 2021, \mnras, 502, 915,
  \dodoi{10.1093/mnras/staa3484}

\bibitem[{{Corbet} {et~al.}(2019){Corbet}, {Chomiuk}, {Coe}, {Coley}, {Dubus},
  {Edwards}, {Martin}, {McBride}, {Stevens}, {Strader}, \& {Townsend}}]{Cor_19}
{Corbet}, R.~H.~D., {Chomiuk}, L., {Coe}, M.~J., {et~al.} 2019, \apj, 884, 93,
  \dodoi{10.3847/1538-4357/ab3e32}

\bibitem[{{Coti Zelati} {et~al.}(2019){Coti Zelati}, {Papitto}, {de Martino},
  {Buckley}, {Odendaal}, {Li}, {Russell}, {Torres}, {Mazzola}, {Bozzo},
  {Gromadzki}, {Campana}, {Rea}, {Ferrigno}, \& {Migliari}}]{Zel19}
{Coti Zelati}, F., {Papitto}, A., {de Martino}, D., {et~al.} 2019, \aap, 622,
  A211, \dodoi{10.1051/0004-6361/201834835}

\bibitem[{{Coti Zelati} {et~al.}(2021){Coti Zelati}, {Hugo}, {Torres}, {de
  Martino}, {Papitto}, {Buckley}, {Russell}, {Campana}, {Van Rooyen}, {Bozzo},
  {Ferrigno}, {Li}, {Migliari}, {Monageng}, {Rea}, {Serylak}, {Stappers}, \&
  {Titus}}]{Zel21}
{Coti Zelati}, F., {Hugo}, B., {Torres}, D.~F., {et~al.} 2021, \aap, 655, A52,
  \dodoi{10.1051/0004-6361/202141431}

\bibitem[{{Desai} {et~al.}(2019){Desai}, {Marchesi}, {Rajagopal}, \&
  {Ajello}}]{Des19}
{Desai}, A., {Marchesi}, S., {Rajagopal}, M., \& {Ajello}, M. 2019, \apjs, 241,
  5, \dodoi{10.3847/1538-4365/ab01fc}

\bibitem[{{Devin} {et~al.}(2020){Devin}, {Lemoine-Goumard}, {Grondin},
  {Castro}, {Ballet}, {Cohen}, \& {Hewitt}}]{G150_Devin20}
{Devin}, J., {Lemoine-Goumard}, M., {Grondin}, M.~H., {et~al.} 2020, \aap, 643,
  A28, \dodoi{10.1051/0004-6361/202038503}

\bibitem[{{Esposito} {et~al.}(2014){Esposito}, {Israel}, {Sidoli}, {Tiengo},
  {Campana}, \& {Moretti}}]{Esp14}
{Esposito}, P., {Israel}, G.~L., {Sidoli}, L., {et~al.} 2014, \mnras, 441,
  1126, \dodoi{10.1093/mnras/stu659}

\bibitem[{{Ferrand} \& {Safi-Harb}(2012)}]{Ferrand2012_SNRCat}
{Ferrand}, G., \& {Safi-Harb}, S. 2012, Advances in Space Research, 49, 1313,
  \dodoi{10.1016/j.asr.2012.02.004}

\bibitem[{{Fortin} {et~al.}(2018){Fortin}, {Chaty}, {Coleiro}, {Tomsick}, \&
  {Nitschelm}}]{For18}
{Fortin}, F., {Chaty}, S., {Coleiro}, A., {Tomsick}, J.~A., \& {Nitschelm},
  C.~H.~R. 2018, \aap, 618, A150, \dodoi{10.1051/0004-6361/201731265}

\bibitem[{{G{\'o}rski} {et~al.}(2005){G{\'o}rski}, {Hivon}, {Banday},
  {Wandelt}, {Hansen}, {Reinecke}, \& {Bartelmann}}]{Gorski2005}
{G{\'o}rski}, K.~M., {Hivon}, E., {Banday}, A.~J., {et~al.} 2005, \apj, 622,
  759, \dodoi{10.1086/427976}

\bibitem[{{Grenier} {et~al.}(2005){Grenier}, {Casandjian}, \&
  {Terrier}}]{EGRET_DarkGas}
{Grenier}, I.~A., {Casandjian}, J.-M., \& {Terrier}, R. 2005, Science, 307,
  1292, \dodoi{10.1126/science.1106924}

\bibitem[{{Grondin} {et~al.}(2011){Grondin}, {Funk}, {Lemoine-Goumard},
  {et~al.}}]{LAT11_J1825}
{Grondin}, M.-H., {Funk}, S., {Lemoine-Goumard}, M., {et~al.} 2011, \apj, 738,
  42, \dodoi{10.1088/0004-637X/738/1/42}

\bibitem[{{Halpern} {et~al.}(2017){Halpern}, {Strader}, \&
  {Li}}]{J0838_Halpern17}
{Halpern}, J.~P., {Strader}, J., \& {Li}, M. 2017, \apj, 844, 150,
  \dodoi{10.3847/1538-4357/aa7cff}

\bibitem[{{Harding} {et~al.}(2005){Harding}, {Usov}, \& {Muslimov}}]{Har05}
{Harding}, A.~K., {Usov}, V.~V., \& {Muslimov}, A.~G. 2005, \apj, 622, 531,
  \dodoi{10.1086/427840}

\bibitem[{{J{\"a}rvel{\"a}} {et~al.}(2020){J{\"a}rvel{\"a}}, {Berton}, {Ciroi},
  {Congiu}, {L{\"a}hteenm{\"a}ki}, \& {Di Mille}}]{Jar_20}
{J{\"a}rvel{\"a}}, E., {Berton}, M., {Ciroi}, S., {et~al.} 2020, \aap, 636,
  L12, \dodoi{10.1051/0004-6361/202037826}

\bibitem[{{Jean} {et~al.}(2018){Jean}, {Cheung}, {Ojha}, {van Zyl}, \&
  {Angioni}}]{Jean_18}
{Jean}, P., {Cheung}, C.~C., {Ojha}, R., {van Zyl}, P., \& {Angioni}, R. 2018,
  ATel, 11546, 1

\bibitem[{{J{\'o}hannesson} \& {Porter}(2021)}]{Gulli2021}
{J{\'o}hannesson}, G., \& {Porter}, T.~A. 2021, \apj, 917, 30,
  \dodoi{10.3847/1538-4357/ac01c9}

\bibitem[{{Johnson} {et~al.}(2018){Johnson}, {Wood}, {Kerr}, {Corbet},
  {Cheung}, {Ray}, \& {Omodei}}]{Assoc18_Joh}
{Johnson}, T.~J., {Wood}, K.~S., {Kerr}, M., {et~al.} 2018, \apj, 863, 27,
  \dodoi{10.3847/1538-4357/aad185}

\bibitem[{{Kennedy} {et~al.}(2020){Kennedy}, {Breton}, {Clark}, {Dhillon},
  {Kerr}, {Buckley}, {Potter}, {S{\'a}nchez}, {Stringer}, \&
  {Marsh}}]{Assoc20_Ken}
{Kennedy}, M.~R., {Breton}, R.~P., {Clark}, C.~J., {et~al.} 2020, \mnras,
  \dodoi{10.1093/mnras/staa912}

\bibitem[{{Kerr}(2010)}]{Kerr2010}
{Kerr}, M. 2010, PhD thesis, University of Washington, ArXiv:1101.6072.
\newblock \doarXiv{1101.6072}

\bibitem[{{Li} {et~al.}(2021){Li}, {Jane Yap}, {Hui}, \& {Kong}}]{Li_21}
{Li}, K.-L., {Jane Yap}, Y.~X., {Hui}, C.~Y., \& {Kong}, A. K.~H. 2021, \apj,
  911, 92, \dodoi{10.3847/1538-4357/abeb76}

\bibitem[{{Li} {et~al.}(2020){Li}, {Kong}, {Aydi}, {Sokolovsky}, {Chomiuk},
  {Kawash}, \& {Strader}}]{Li20}
{Li}, K.-L., {Kong}, A., {Aydi}, E., {et~al.} 2020, ATel, 13868, 1

\bibitem[{{Li} {et~al.}(2016){Li}, {Kong}, {Hou}, {Mao}, {Strader}, {Chomiuk},
  \& {Tremou}}]{J0212_Li16}
{Li}, K.-L., {Kong}, A. K.~H., {Hou}, X., {et~al.} 2016, \apj, 833, 143,
  \dodoi{10.3847/1538-4357/833/2/143}

\bibitem[{{Manchester} {et~al.}(2005){Manchester}, {Hobbs}, {Teoh}, \&
  {Hobbs}}]{ATNFcatalog}
{Manchester}, R.~N., {Hobbs}, G.~B., {Teoh}, A., \& {Hobbs}, M. 2005, \aj, 129,
  1993, \dodoi{10.1086/428488}

\bibitem[{{Mares} {et~al.}(2021){Mares}, {Lemoine-Goumard}, {Acero}, {Clark},
  {Devin}, {Gabici}, {Gelfand}, {Green}, \& {Grondin}}]{HESSJ1640_Mares21}
{Mares}, A., {Lemoine-Goumard}, M., {Acero}, F., {et~al.} 2021, \apj, 912, 158,
  \dodoi{10.3847/1538-4357/abef62}

\bibitem[{{Mart{\'\i}} {et~al.}(2020){Mart{\'\i}}, {S{\'a}nchez-Ayaso},
  {Luque-Escamilla}, {Paredes}, {Bosch-Ramon}, \& {Corbet}}]{Assoc20_Marti}
{Mart{\'\i}}, J., {S{\'a}nchez-Ayaso}, E., {Luque-Escamilla}, P.~L., {et~al.}
  2020, \mnras, 492, 4291, \dodoi{10.1093/mnras/staa072}

\bibitem[{{Mart{\'\i}-Devesa} \& {Reimer}(2020)}]{Mar_20}
{Mart{\'\i}-Devesa}, G., \& {Reimer}, O. 2020, \aap, 637, A23,
  \dodoi{10.1051/0004-6361/202037442}

\bibitem[{{Mart{\'\i}-Devesa} {et~al.}(2020){Mart{\'\i}-Devesa}, {Reimer},
  {Li}, \& {Torres}}]{Marti_20}
{Mart{\'\i}-Devesa}, G., {Reimer}, O., {Li}, J., \& {Torres}, D.~F. 2020, \aap,
  635, A141, \dodoi{10.1051/0004-6361/202037462}

\bibitem[{{McConville} {et~al.}(2011){McConville}, {Ostorero}, {Moderski},
  {Stawarz}, {Cheung}, {Ajello}, {Bouvier}, {Bregeon}, {Donato}, {Finke},
  {Furniss}, {McEnery}, {Monzani}, {Orienti}, {Reyes}, {Rossetti}, \&
  {Williams}}]{LAT11_4C55}
{McConville}, W., {Ostorero}, L., {Moderski}, R., {et~al.} 2011, \apj, 738,
  148, \dodoi{10.1088/0004-637X/738/2/148}

\bibitem[{Miller {et~al.}(2020)Miller, Swihart, Strader, Urquhart, Aydi,
  Chomiuk, Dage, Kawash, Shishkovsky, \& Sokolovsky}]{Mil20}
Miller, J.~M., Swihart, S.~J., Strader, J., {et~al.} 2020, The Astrophysical
  Journal, 904, 49, \dodoi{10.3847/1538-4357/abbb2e}

\bibitem[{{NASA/IPAC Extragalactic Database (NED)}(2019)}]{NED1}
{NASA/IPAC Extragalactic Database (NED)}. 2019, NASA/IPAC Extragalactic
  Database (NED),  IPAC, \dodoi{10.26132/NED1}

\bibitem[{{Paliya}(2021)}]{Pal21}
{Paliya}, V.~S. 2021, \apjl, 918, L39, \dodoi{10.3847/2041-8213/ac2143}

\bibitem[{{Pe{\~n}a-Herazo} {et~al.}(2017){Pe{\~n}a-Herazo}, {Marchesini},
  {{\'A}lvarez Crespo}, {Ricci}, {Massaro}, {Chavushyan}, {Landoni}, {Strader},
  {Chomiuk}, {Cheung}, {Masetti}, {Jim{\'e}nez-Bail{\'o}n}, {D'Abrusco},
  {Paggi}, {Milisavljevic}, {La Franca}, {Smith}, \& {Tosti}}]{Fup_Pen17}
{Pe{\~n}a-Herazo}, H.~A., {Marchesini}, E.~J., {{\'A}lvarez Crespo}, N.,
  {et~al.} 2017, \apss, 362, 228, \dodoi{10.1007/s10509-017-3208-7}

\bibitem[{{Pe{\~n}a-Herazo} {et~al.}(2019){Pe{\~n}a-Herazo}, {Massaro},
  {Chavushyan}, {Marchesini}, {Paggi}, {Landoni}, {Masetti}, {Ricci},
  {D'Abrusco}, {Milisavljevic}, {Jim{\'e}nez-Bail{\'o}n}, {La Franca}, {Smith},
  \& {Tosti}}]{Fup_Pen19}
{Pe{\~n}a-Herazo}, H.~A., {Massaro}, F., {Chavushyan}, V., {et~al.} 2019,
  \apss, 364, 85, \dodoi{10.1007/s10509-019-3574-4}

\bibitem[{{Pe{\~n}a-Herazo} {et~al.}(2020){Pe{\~n}a-Herazo},
  {Amaya-Almaz{\'a}n}, {Massaro}, {de Menezes}, {Marchesini}, {Chavushyan},
  {Paggi}, {Landoni}, {Masetti}, {Ricci}, {D'Abrusco}, {Cheung}, {La Franca},
  {Smith}, {Milisavljevic}, {Jim{\'e}nez-Bail{\'o}n}, {Pati{\~n}o-{\'A}lvarez},
  \& {Tosti}}]{Pen20}
{Pe{\~n}a-Herazo}, H.~A., {Amaya-Almaz{\'a}n}, R.~A., {Massaro}, F., {et~al.}
  2020, \aap, 643, A103, \dodoi{10.1051/0004-6361/202037978}

\bibitem[{{Pe{\~n}a-Herazo} {et~al.}(2021{\natexlab{a}}){Pe{\~n}a-Herazo},
  {Paggi}, {Garc{\'\i}a-P{\'e}rez}, {Amaya-Almaz{\'a}n}, {Massaro}, {Ricci},
  {Chavushyan}, {Marchesini}, {Masetti}, {Landoni}, {D'Abrusco},
  {Milisavljevic}, {Jim{\'e}nez-Bail{\'o}n}, {Pati{\~n}o-{\'A}lvarez}, {La
  Franca}, {Smith}, \& {Tosti}}]{Pen21a}
{Pe{\~n}a-Herazo}, H.~A., {Paggi}, A., {Garc{\'\i}a-P{\'e}rez}, A., {et~al.}
  2021{\natexlab{a}}, \aj, 162, 177, \dodoi{10.3847/1538-3881/ac1da7}

\bibitem[{{Pe{\~n}a-Herazo} {et~al.}(2021{\natexlab{b}}){Pe{\~n}a-Herazo},
  {Massaro}, {Gu}, {Paggi}, {Landoni}, {D'Abrusco}, {Ricci}, {Masetti}, \&
  {Chavushyan}}]{Pen21b}
{Pe{\~n}a-Herazo}, H.~A., {Massaro}, F., {Gu}, M., {et~al.} 2021{\natexlab{b}},
  \aj, 162, 76, \dodoi{10.3847/1538-3881/ac09e2}

\bibitem[{{Pe{\~n}a-Herazo} {et~al.}(2021{\natexlab{c}}){Pe{\~n}a-Herazo},
  {Massaro}, {Gu}, {Paggi}, {Landoni}, {D'Abrusco}, {Ricci}, {Masetti}, \&
  {Chavushyan}}]{Pen21c}
---. 2021{\natexlab{c}}, \aj, 161, 196, \dodoi{10.3847/1538-3881/abe41d}

\bibitem[{{Porter} {et~al.}(2017){Porter}, {J{\'o}hannesson}, \&
  {Moskalenko}}]{GALPROP17}
{Porter}, T.~A., {J{\'o}hannesson}, G., \& {Moskalenko}, I.~V. 2017, \apj, 846,
  67, \dodoi{10.3847/1538-4357/aa844d}

\bibitem[{{Principe} {et~al.}(2020){Principe}, {Mitchell}, {Caroff}, {Hinton},
  {Parsons}, \& {Funk}}]{HESSJ1825_Principe20}
{Principe}, G., {Mitchell}, A.~M.~W., {Caroff}, S., {et~al.} 2020, \aap, 640,
  A76, \dodoi{10.1051/0004-6361/202038375}

\bibitem[{{Rajagopal} {et~al.}(2021){Rajagopal}, {Marchesi}, {Kaur},
  {Dom{\'\i}nguez}, {Silver}, \& {Ajello}}]{Raj21}
{Rajagopal}, M., {Marchesi}, S., {Kaur}, A., {et~al.} 2021, \apjs, 254, 26,
  \dodoi{10.3847/1538-4365/abf656}

\bibitem[{{Selig} {et~al.}(2015){Selig}, {Vacca}, {Oppermann}, \&
  {En{\ss}lin}}]{D3PO15}
{Selig}, M., {Vacca}, V., {Oppermann}, N., \& {En{\ss}lin}, T.~A. 2015, \aap,
  581, A126, \dodoi{10.1051/0004-6361/201425172}

\bibitem[{{Smith} {et~al.}(2019){Smith}, {Bruel}, {Cognard}, {Cameron},
  {Camilo}, {Dai}, {Guillemot}, {Johnson}, {Johnston}, {Keith}, {Kerr},
  {Kramer}, {Lyne}, {Manchester}, {Shannon}, {Sobey}, {Stappers}, \&
  {Weltevrede}}]{Smi19}
{Smith}, D.~A., {Bruel}, P., {Cognard}, I., {et~al.} 2019, \apj, 871, 78,
  \dodoi{10.3847/1538-4357/aaf57d}

\bibitem[{{Strader} {et~al.}(2014){Strader}, {Chomiuk}, {Sonbas}, {Sokolovsky},
  {Sand}, {Moskvitin}, \& {Cheung}}]{J0523_Strader14}
{Strader}, J., {Chomiuk}, L., {Sonbas}, E., {et~al.} 2014, \apjl, 788, L27,
  \dodoi{10.1088/2041-8205/788/2/L27}

\bibitem[{{Strader} {et~al.}(2021){Strader}, {Swihart}, {Urquhart}, {Chomiuk},
  {Aydi}, {Bahramian}, {Kawash}, {Sokolovsky}, {Tremou}, \& {Udalski}}]{Stra21}
{Strader}, J., {Swihart}, S.~J., {Urquhart}, R., {et~al.} 2021, \apj, 917, 69,
  \dodoi{10.3847/1538-4357/ac0b47}

\bibitem[{{Swihart} {et~al.}(2021){Swihart}, {Strader}, {Aydi}, {Chomiuk},
  {Dage}, \& {Shishkovsky}}]{Swi_21}
{Swihart}, S.~J., {Strader}, J., {Aydi}, E., {et~al.} 2021, \apj, 909, 185,
  \dodoi{10.3847/1538-4357/abe1be}

\bibitem[{{Tabassum} {et~al.}(2021){Tabassum}, {Ransom}, {Ray}, {Cromartie},
  {Al Noori}, {Camilo}, {Roberts}, {Ferrara}, \& {Clark}}]{Tab21}
{Tabassum}, S., {Ransom}, S., {Ray}, P., {et~al.} 2021, in 43rd COSPAR
  Scientific Assembly. Held 28 January - 4 February, Vol.~43, 1208

\bibitem[{{Taylor}(2005)}]{Tay05}
{Taylor}, M.~B. 2005, in Astronomical Society of the Pacific Conference Series,
  Vol. 347, Astronomical Data Analysis Software and Systems XIV, ed.
  P.~{Shopbell}, M.~{Britton}, \& R.~{Ebert}, 29

\bibitem[{{Tibaldo} {et~al.}(2018){Tibaldo}, {Zanin}, {Faggioli}, {Ballet},
  {Grondin}, {Hinton}, \& {Lemoine-Goumard}}]{VelaX_Tibaldo18}
{Tibaldo}, L., {Zanin}, R., {Faggioli}, G., {et~al.} 2018, \aap, 617, A78,
  \dodoi{10.1051/0004-6361/201833356}

\bibitem[{{Venter} {et~al.}(2009){Venter}, {De Jager}, \& {Clapson}}]{Ven09}
{Venter}, C., {De Jager}, O.~C., \& {Clapson}, A.~C. 2009, \apjl, 696, L52,
  \dodoi{10.1088/0004-637X/696/1/L52}

\bibitem[{{Wang} {et~al.}(2020){Wang}, {Xing}, {Zhang}, {Boutsia}, {Wang}, {V},
  {Burdge}, {Coughlin}, {Duev}, {Kulkarni}, {Riddle}, \&
  {Serabyn}}]{Assoc20_Wang}
{Wang}, Z., {Xing}, Y., {Zhang}, J., {et~al.} 2020, \mnras, 493, 4845,
  \dodoi{10.1093/mnras/staa655}

\bibitem[{{Xin} {et~al.}(2019){Xin}, {Zeng}, {Liu}, {Fan}, \&
  {Wei}}]{VERJ2227_Xin19}
{Xin}, Y., {Zeng}, H., {Liu}, S., {Fan}, Y., \& {Wei}, D. 2019, \apj, 885, 162,
  \dodoi{10.3847/1538-4357/ab48ee}

\end{thebibliography}
 
\appendix

\section{Description of the FITS version of the 4FGL-DR3 catalog}
\label{appendix_fits_format}

\startlongtable
\begin{deluxetable*}{lccl}
\setlength{\tabcolsep}{0.04in}
\tablewidth{0pt}
\tabletypesize{\scriptsize}
\tablecaption{LAT 4FGL-DR3 FITS format: LAT\_Point\_Source\_Catalog extension. Changes since 4FGL appear in $Italics$. The file is available at \url{https://fermi.gsfc.nasa.gov/ssc/data/access/lat/12yr_catalog/}.
\label{tab:description}}
\tablehead{
\colhead{Column} &
\colhead{Format} &
\colhead{Unit} &
\colhead{Description}
}
\startdata
Source\_Name & 18A & \nodata & Source name 4FGL JHHMM.m+DDMMa\tablenotemark{a} \\
$DataRelease$ & I & \nodata & 1 for DR1, 2 for new in DR2, 3 for new or changed in DR3 \\
RAJ2000 & E & deg & Right Ascension \\
DEJ2000 & E & deg & Declination \\
GLON & E & deg & Galactic longitude \\
GLAT & E & deg & Galactic latitude \\
Conf\_68\_SemiMajor & E & deg & Long radius of error ellipse at 68\% confidence\tablenotemark{b} \\
Conf\_68\_SemiMinor & E & deg & Short radius of error ellipse at 68\% confidence\tablenotemark{b} \\
Conf\_68\_PosAng & E & deg & Position angle of the 68\% ellipse\tablenotemark{b} \\
Conf\_95\_SemiMajor & E & deg & Long radius of error ellipse at 95\% confidence \\
Conf\_95\_SemiMinor & E & deg & Short radius of error ellipse at 95\% confidence \\
Conf\_95\_PosAng & E & deg & Position angle (eastward) of the long axis from celestial north \\
ROI\_num & I & \nodata & RoI number (cross-reference to ROIs extension) \\
Extended\_Source\_Name & 18A & \nodata & Cross-reference to the ExtendedSources extension \\
Signif\_Avg & E & \nodata & Source significance in $\sigma$ units over the 100~MeV to 1~TeV band \\
Pivot\_Energy & E & MeV & Energy at which error on differential flux is minimal \\
Flux1000 & E & cm$^{-2}$ s$^{-1}$ & Integral photon flux from 1 to 100~GeV \\
Unc\_Flux1000 & E & cm$^{-2}$ s$^{-1}$ & $1\sigma$ error on integral photon flux from 1 to 100~GeV \\
Energy\_Flux100 & E & erg cm$^{-2}$ s$^{-1}$ & Energy flux from 100~MeV to 100~GeV obtained by spectral fitting \\
Unc\_Energy\_Flux100 & E & erg cm$^{-2}$ s$^{-1}$ & $1\sigma$  error on energy flux from 100~MeV to 100~GeV \\
SpectrumType & 18A & \nodata & Spectral type in the global model (PowerLaw, LogParabola, PLSuperExpCutoff) \\
PL\_Flux\_Density & E & cm$^{-2}$ MeV$^{-1}$ s$^{-1}$ & Differential flux at Pivot\_Energy in PowerLaw fit \\
Unc\_PL\_Flux\_Density & E & cm$^{-2}$ MeV$^{-1}$ s$^{-1}$ & $1\sigma$  error on PL\_Flux\_Density \\
PL\_Index & E & \nodata & Photon index when fitting with PowerLaw \\
Unc\_PL\_Index & E & \nodata & $1\sigma$ error on PL\_Index \\
LP\_Flux\_Density & E & cm$^{-2}$ MeV$^{-1}$ s$^{-1}$ & Differential flux at Pivot\_Energy in LogParabola fit \\
Unc\_LP\_Flux\_Density & E & cm$^{-2}$ MeV$^{-1}$ s$^{-1}$ & $1\sigma$  error on LP\_Flux\_Density \\
LP\_Index & E & \nodata & Photon index at Pivot\_Energy ($\alpha$ of Eq.~\ref{eq:logparabola}) when fitting with LogParabola \\
Unc\_LP\_Index & E & \nodata & $1\sigma$ error on LP\_Index \\
LP\_beta & E & \nodata & Curvature parameter ($\beta$ of Eq.~\ref{eq:logparabola}) when fitting with LogParabola \\
Unc\_LP\_beta & E & \nodata & $1\sigma$ error on LP\_beta \\
LP\_SigCurv & E & \nodata & Significance (in $\sigma$ units) of the fit improvement between PowerLaw and \\
& & & LogParabola. A value greater than 4 indicates significant curvature \\
$LP\_EPeak$ & E & MeV & Peak energy in $\nu F_\nu$ estimated from the LogParabola model \\
$Unc\_LP\_EPeak$ & E & MeV & $1\sigma$ error on LP\_EPeak \\
PLEC\_Flux\_Density & E & cm$^{-2}$ MeV$^{-1}$ s$^{-1}$ & Differential flux at Pivot\_Energy in PLSuperExpCutoff fit \\
Unc\_PLEC\_Flux\_Density & E & cm$^{-2}$ MeV$^{-1}$ s$^{-1}$ & $1\sigma$  error on PLEC\_Flux\_Density \\
$PLEC\_IndexS$ & E & \nodata & Photon index at Pivot\_Energy ($\Gamma_S$ of Eq.~\ref{eq:expcutoff}) when fitting with \\
& & & PLSuperExpCutoff \\
$Unc\_PLEC\_IndexS$ & E & \nodata & $1\sigma$ error on PLEC\_IndexS \\
$PLEC\_ExpfactorS$ & E & \nodata & Spectral curvature at Pivot\_Energy ($d$ of Eq.~\ref{eq:expcutoff}) when fitting with \\
& & & PLSuperExpCutoff \\
$Unc\_PLEC\_ExpfactorS$ & E & \nodata & $1\sigma$ error on PLEC\_ExpfactorS \\
PLEC\_Exp\_Index & E & \nodata & Exponential index ($b$ of Eq.~\ref{eq:expcutoff}) when fitting with PLSuperExpCutoff \\
Unc\_PLEC\_Exp\_Index & E & \nodata & $1\sigma$ error on PLEC\_Exp\_Index \\
PLEC\_SigCurv & E & \nodata & Same as LP\_SigCurv for PLSuperExpCutoff model \\
$PLEC\_EPeak$ & E & MeV & Peak energy in $\nu F_\nu$ estimated from the PLSuperExpCutoff model \\
$Unc\_PLEC\_EPeak$ & E & MeV & $1\sigma$ error on PLEC\_EPeak \\
Npred & E & \nodata & Predicted number of events in the model \\
Flux\_Band & 8E & cm$^{-2}$ s$^{-1}$ & Integral photon flux in each spectral band \\
Unc\_Flux\_Band & $2 \times 8$E & cm$^{-2}$ s$^{-1}$ & $1\sigma$ lower and upper error on Flux\_Band\tablenotemark{c} \\
nuFnu\_Band & 8E & erg cm$^{-2}$ s$^{-1}$ & Spectral energy distribution over each spectral band \\
Sqrt\_TS\_Band & 8E & \nodata & Square root of the Test Statistic in each spectral band \\
Variability\_Index & E & \nodata & Sum of 2$\times$log(Likelihood) difference between the flux fitted in each time \\
& & & interval and the average flux over the full catalog interval; a value greater \\
& & & than 24.72 over 12 intervals indicates $< $1\% chance of being a steady source \\
Frac\_Variability & E & \nodata & Fractional variability computed from the fluxes in each year \\
Unc\_Frac\_Variability & E & \nodata & $1\sigma$ error on fractional variability \\
Signif\_Peak & E & \nodata & Source significance in peak interval in $\sigma$ units \\
Flux\_Peak & E & cm$^{-2}$ s$^{-1}$ & Peak integral photon flux from 100~MeV to 100~GeV \\
Unc\_Flux\_Peak & E & cm$^{-2}$ s$^{-1}$ &  $1\sigma$ error on peak integral photon flux \\
Time\_Peak & D & s (MET) & Time of center of interval in which peak flux was measured \\
Peak\_Interval & E & s & Length of interval in which peak flux was measured \\
Flux\_History & 12E & cm$^{-2}$ s$^{-1}$ & Integral photon flux from 100~MeV to 100~GeV in each year (best fit from \\
& & &  likelihood analysis with spectral shape fixed to that obtained over full interval)\\
Unc\_Flux\_History & $2 \times 12$E & cm$^{-2}$ s$^{-1}$ &  $1\sigma$ lower and upper error on integral photon flux in each year\tablenotemark{c} \\
Sqrt\_TS\_History & 12E & \nodata & Square root of the Test Statistic in each year \\
$ASSOC\_4FGL$ & 18A & \nodata & Correspondence to 4FGL source catalog, if any \\
ASSOC\_FGL & 18A & \nodata & Most recent correspondence to FGL source catalogs prior to 4FGL, if any \\
ASSOC\_FHL & 18A & \nodata & Most recent correspondence to previous FHL source catalogs, if any \\
ASSOC\_GAM1 & 18A & \nodata & Name of likely corresponding 2AGL source, if any \\
ASSOC\_GAM2 & 18A & \nodata & Name of likely corresponding 3EG source, if any \\
ASSOC\_GAM3 & 18A & \nodata & Name of likely corresponding EGR source, if any \\
TEVCAT\_FLAG & A & \nodata & P if positional association with nonextended source in TeVCat \\
& & & E if associated with an extended source in TeVCat, N if no TeV association \\
ASSOC\_TEV & 24A & \nodata & Name of likely corresponding TeV source from TeVCat, if any \\
CLASS1 & 5A & \nodata & Class designation for associated source; see Table~\ref{tab:classes} \\
CLASS2 & 5A & \nodata & Class designation for low-confidence association \\
ASSOC1 & 28A & \nodata & Name of identified or likely associated source \\
ASSOC2 & 28A & \nodata & Name of low-confidence association or of enclosing extended source \\
ASSOC\_PROB\_BAY & E & \nodata & Probability of association according to the Bayesian method\tablenotemark{d} \\
ASSOC\_PROB\_LR & E & \nodata & Probability of association according to the Likelihood Ratio method\tablenotemark{e} \\
RA\_Counterpart & D & deg & R.A. of the counterpart ASSOC1 \\
DEC\_Counterpart & D & deg & Decl. of the counterpart ASSOC1 \\
Unc\_Counterpart & E & deg & 95\% precision of the counterpart localization\tablenotemark{f} \\
Flags & I & \nodata & Source flags (binary coding as in Table~\ref{tab:flags})\tablenotemark{g} \\
\enddata
\tablenotetext{a} {The coordinates are rounded, following the International Astronomical Union convention. The letter at the end can be \texttt{c} (coincident with interstellar clump), 
\texttt{e} (extended source), \texttt{i} (for Crab nebula inverse Compton) or \texttt{s} (for Crab nebula synchrotron).}
\tablenotetext{b} {From the 95\% ellipse, assuming a Gaussian distribution.}
\tablenotetext{c} {Separate $1\sigma$ errors are computed from the likelihood profile toward lower and larger fluxes. The lower error is set equal to NULL, and the upper error is derived from a Bayesian upper limit if the $1\sigma$ interval contains 0 ($TS < 1$).}
\tablenotetext{d} {NaN in this column when ASSOC1 is defined means that the probability could not be computed, either because the source is extended or because the counterpart is the result of dedicated follow-up.}
\tablenotetext{e} {Probabilities $<$ 0.8 are formally set to 0.}
\tablenotetext{f} {For extended counterparts, this reports their extension radius.}
\tablenotetext{g} {Each condition is indicated by one bit among the 16 bits forming \texttt{Flags}. The bit is raised (set to 1) in the dubious case, so that sources without any warning sign have \texttt{Flags} = 0.}
\end{deluxetable*}

The FITS format version of the third release of the 4FGL catalog has seven binary table extensions.  The extension {\tt LAT\_Point\_Source\_Catalog Extension} has all of the information about the sources, and has changed to adapt to the evolutions described in \S~\ref{catalog_spectra}. Its format is described in Table~\ref{tab:description}. Columns \texttt{(Unc\_)PLEC\_Index} and \texttt{(Unc\_)PLEC\_Expfactor} have been replaced by \texttt{(Unc\_)PLEC\_IndexS} and \texttt{(Unc\_)PLEC\_ExpfactorS} corresponding to the new pulsar parameterization (Eq.~\ref{eq:expcutoff}). The table has 6659 rows for 6658 sources because the Crab nebula is described by two entries (the synchrotron component and the inverse Compton component) but counted as only one source. The Crab pulsar is another entry and counted as a separate source. The extended sources are indicated by an ``e'' appended to their names in the main table.

The extensions {\tt GTI}, {\tt ExtendedSources}, {\tt ROIs}, {\tt Components}, {\tt EnergyBounds}, {\tt Hist\_Start} describe auxiliary details and have the same format as in 4FGL.

\end{document}